\documentclass[12pt]{JHEP3}

\usepackage[dvips]{epsfig}

 \newlength{\wth}
\newcommand{\twographst}[2]{%
 \unitlength=1.1in
 \begin{picture}(5.8,2.3)(0.5,0.25)
 \put(-0.04,2.54){\epsfig{file=#1, width=0.698 \wth,angle=270}}
 \put(0.85,0.5){\epsfig{file=#12, width=0.68 \wth}}
 \put(2.66,2.54){\epsfig{file=#2, width=0.698 \wth, angle=270}}
 \put(3.55,0.5){\epsfig{file=#22, width=0.68 \wth}}
 \put(0.5,2.1){(a)}
 \put(3.2,2.1){(b)}
 \end{picture}
}

\newcommand{\fourgraphst}[4]{%
 \unitlength=1.1in
 \begin{picture}(5.8,4)(0.5,0.4)
\put(0,2){\put(-0.04,2.54){\epsfig{file=#1, width=0.698 \wth,angle=270}}
 \put(0.85,0.5){\epsfig{file=#12, width=0.68 \wth}}
 \put(2.66,2.54){\epsfig{file=#2, width=0.698 \wth, angle=270}}
 \put(3.55,0.5){\epsfig{file=#22, width=0.68 \wth}}
 \put(0.5,2.1){(a)}
 \put(3.2,2.1){(b)}
}

\put(0,0){\put(-0.04,2.54){\epsfig{file=#3, width=0.698 \wth,angle=270}}
 \put(0.85,0.5){\epsfig{file=#32, width=0.68 \wth}}
 \put(2.66,2.54){\epsfig{file=#4, width=0.698 \wth, angle=270}}
 \put(3.55,0.5){\epsfig{file=#42, width=0.68 \wth}}
 \put(0.5,2.1){(c)}
 \put(3.2,2.1){(d)}
}
 \end{picture}
}

\newcommand{\fourgraphs}[4]{%
 \unitlength=1.1in
 \begin{picture}(5.8,4.4)(0.3,0.3)
\put(0,2.4){\put(0.5,0){\epsfig{file=#1, width=0.698 \wth}}
 \put(2.9,0){\epsfig{file=#2, width=0.698 \wth}}
 \put(0.5,2.2){(a)}
 \put(3,2.2){(b)}}
\put(0,0){\put(0.5,0){\epsfig{file=#3, width=0.698 \wth}}
 \put(2.9,0){\epsfig{file=#4, width=0.698 \wth}}
 \put(0.5,2.2){(c)}
 \put(3,2.2){(d)}}
 \end{picture}
}

\newcommand{\onegraph}[1]{%
 \unitlength=1.1in
 \begin{picture}(2.5,2.3)(0.5,0.25)
 \put(-0.04,2.54){\epsfig{file=#1, width=0.698 \wth,angle=270}}
 \put(0.85,0.5){\epsfig{file=#12, width=0.68 \wth}}
 \end{picture}
}

 \newcommand{\twographs}[2]{%
 \unitlength=1.1in
 \begin{picture}(5.8,2.3)(0,0.25)
 \put(0,0){\epsfig{file=#1.eps, width=0.698 \wth}}
 \put(2.7,0){\epsfig{file=#2.eps, width=0.698 \wth}}
 \put(0.0,2.3){(a)}
 \put(2.7,2.3){(b)}
 \end{picture}
}

\title{The Dark Side of mSUGRA}

\author{Benjamin C Allanach$^{1}$, Christopher G Lester$^{2}$ and Arne M
  Weber$^{3}$ \\ 
$^{1}$ DAMTP, CMS, Wilberforce Road, Cambridge CB3 0WA, UK\\
$^{2}$ Cavendish Laboratory, J.J. Thomson Avenue, Cambridge CB3 0HE, UK\\
$^3$ Max Planck Inst.\ f\"{u}r Phys., F\"{o}hringer Ring 6, D-80805 Munich,
  Germany\\ 
}

\abstract{We study the $\mu<0$ branch of the minimal supergravity
  ansatz of the minimal supersymmetric standard model. The extent to
  which $\mu<0$ is disfavoured compared to $\mu>0$ in global fits is
  calculated 
  with Markov Chain Monte Carlo methods and bridge sampling. The fits
  include   state-of-the-art two-loop MSSM contributions to the electroweak
  observables 
  $M_W$ and $\sin^2 \theta_w^l$, as well as the anomalous magnetic moment of the
  muon $(g-2)_\mu$, the relic density of dark matter and other
  relevant indirect observables. $\mu<0$ is only
  marginally disfavoured in global fits and should be considered in mSUGRA
  analyses. We estimate that the ratio of 
  probabilities is $P(\mu<0) / P(\mu>0)=0.07-0.16$. 
}

\preprint{DAMTP-2006-75\\MPP--2006--122\\ Cavendish-HEP-2006-024 \\
  hep-ph/0609295} 

\begin{document}

\section{Introduction}

There has been increasing recent attention on global fits of various indirect
data to minimal supergravity (mSUGRA), also sometimes called the constrained
minimal supersymmetric standard model
(CMSSM)~\cite{Ellis:2003si,Profumo:2004at,Baltz:2004aw,Ellis:2004tc,Stark:2005mp}.
mSUGRA makes phenomenological 
analysis of the minimal supersymmetric standard model (MSSM)
tractable via the low number of free parameters.
In fact, the scalar masses $m_0$, 
gaugino masses $M_{1/2}$ and trilinear coupling $A_0$ are assumed to be
universal at a gauge unification scale  $M_{GUT} \sim 2 \times 10^{16}$ GeV. 
If the MSSM is present in nature and if the mSUGRA
universality assumptions are approximately correct, chi-squared or probability
distributions for potential collider/dark matter observables
can be derived. Early
fits~\cite{Robros,Ellis:2003si,Profumo:2004at,Baltz:2004aw,Ellis:2004tc}
necessarily had fixed input parameters to reduce 
the dimensionality of the input parameter space, making scans practicable.
It is usually assumed that neutralinos constitute the current cold dark matter
content of the universe, since they are weakly interacting, electrically and
colour neutral and stable.
The predicted value of dark matter relic density $\Omega_{DM} h^2$ is a very
strong constraint on viable mSUGRA parameter space, effectively reducing its
dimensionality by 1.

The accuracy of the inferred value of $\Omega_{DM} h^2$ from WMAP data makes a
global fit to all of the relevant mSUGRA parameters potentially difficult
because the system is rather under-constrained, possessing narrow, steep
valleys of degenerate $\chi^2$ minima.  
If the MSSM is confirmed in colliders, it will hopefully be possible to break
such degeneracies with collider observables. This does not help us at present,
where we want to provide a sort of `weather forecast' for future colliders.
It was indicated in ref.~\cite{Baltz:2004aw} that the powerful Markov Chain
Monte Carlo (MCMC) technique might allow us to find the probability
distribution of a  fully global fit to indirect data. 
Two of us went on~\cite{Allanach:2005kz} to demonstrate that MCMCs do indeed
allow such a fit, investigating collider observables. One of us examined the
effect of a naturalness prior~\cite{Allanach:2006jc}. Our results were
confirmed and expanded in Ref.~\cite{deAustri:2006pe}, also utilising the MCMC
method and including a one-loop MSSM calculation of the W-boson mass $M_W$ and 
the weak leptonic mixing angle $\sin^2 \theta_w^l$ in the likelihood density.
The purpose of MCMC mSUGRA global fits is two-fold: as well as producing
interesting and useful physics results in themselves, we may profit from the
experience of utilising and developing the MCMC tools, which could prove very
useful when analysing future collider data. 

It is our purpose in the present paper to extend the previous $\mu>0$ global
fits to $\mu<0$. 
Besides the observables studied in~\cite{Allanach:2005kz} we now also include
$M_W$ and $\sin \theta_w^l$ in our analysis, as done in
e.g. \cite{Ellis:2004tc,deAustri:2006pe}. Being highly sensitive to new physics
these very accurately measured quantities play a key role in the electroweak
sector and are therefore also of great interest when it comes to further
constraining the mSUGRA parameter space. It was shown in the literature and
that the one-loop predictions for
the two observables alone do not bear enough accuracy to make reliable
predictions. In fact, the pure one-loop predictions can lead to results
contradictory to the state-of-the-art predictions \cite{Heinemeyer:2006px}
used in our analysis. These contain the known higher order contributions from
both the Standard Model and the MSSM\@. Extending our analysis to negative
values of 
$\mu$ it is crucial to further use a very accurate prediction for
$(g-2)_\mu$. The dominant two-loop corrections \cite{2loop} to this quantity
are therefore also taken into account in the present analysis. 
It is well known that the measured anomalous value of the magnetic
moment of the muon $(g-2)_\mu$ is roughly 2$\sigma$ above the Standard
Model (SM) predicted value. This positive contribution is predicted by
some regions of mSUGRA parameter space, provided $\mu>0$. The ``dark
side'' of mSUGRA (i.e.\ $\mu<0$) provides a negative contribution,
thereby being disfavoured by the $(g-2)_\mu$ measurement. In a global
fit, one can trade likelihood penalties between different observables
and the conclusion that $\mu<0$ is disfavoured to roughly 2$\sigma$ is
not at all obvious. We will calculate the extent to which the dark
side is ruled out by using MCMCs with ``bridge sampling''
\cite{radford}.  
We will encounter
problems associated with isolated likelihood density maxima in the
dark side, potentially ruining MCMC convergence. Fortunately, bridge
sampling provides a solution to the convergence issue and we are able
to calculate the degree to which the dark side is disfavoured with
respect to $\mu>0$. 
As well as extending
previous analyses to $\mu<0$, we have made several technical
improvements in the calculation of the likelihood compared with previous
attempts in the literature.

If the lightest supersymmetric particle decays into Standard Model
particles, as is the case in R-parity violation for instance, its
relic density will be essentially zero today. In that case, one
requires to obtain the WMAP fitted $\Omega_{DM} h^2$ from some other
source than neutralinos (gravitinos or hidden sector matter for
instance).  In order to investigate this case, we will also perform
the fits for the case where all relevant data {\em except}
$\Omega_{DM} h^2$ are included in the likelihood density.  Such fits
will help us to understand the impact of $\Omega_{DM} h^2$ in
constraining the model, as well as being relevant for the R-parity
violating mSUGRA~\cite{rpvSUGRA} in the limit of small R-parity
violating couplings.

We now go on to detail the various constraints used on the model in
section~\ref{sec:const}. The results of the dark side fitting
procedure are compared and contrasted against the better known $\mu>0$ ones
in section~\ref{sec:dark}, before the effect of dropping the dark-matter
constraint is examined in section~\ref{sec:noDM}. 
Closing remarks are presented in section~\ref{sec:conc}. A presentation of the
fitting procedures is confined to the appendix: 
Markov Chain Monte Carlos and bridge sampling are discussed in
appendix~\ref{sec:bridge}. Convergence problems
and their resolution are discussed in appendix~\ref{sec:conv}. 

\section{Constraints \label{sec:const}}

\TABULAR[r]{|c|c|}{\hline
mSUGRA parameter & range \\ \hline
$A_0$ & -4 TeV to 4 TeV \\
$m_0$ & 60 GeV to 4 TeV \\
$M_{1/2}$ & 60 GeV to 2 TeV \\
$\tan \beta$ &  2 to 62 \\ \hline
SM parameter & constraint \\ \hline
$1/\alpha^{\overline{MS}}$ & 127.918$\pm$0.018 \\
$\alpha_s^{\overline{MS}}(M_Z)$ & 0.1176$\pm$0.002 \\
$m_b(m_b)^{\overline   MS}$ & 4.24$\pm$0.11 GeV \\
$m_t$ & 171.4$\pm$2.1 \\ \hline
}{Input parameters \label{tab:inp}}
We vary 8 input parameters relevant to the model. 
The range of intrinsically mSUGRA parameters considered is shown in
Table~\ref{tab:inp}, where $\tan \beta$ is the ratio of the two MSSM Higgs
doublet vacuum expectation values.
We use Ref.~\cite{PDG} for the QED coupling
constant $\alpha^{\overline{MS}}$, the strong coupling constant
$\alpha_s^{\overline{MS}}(M_Z)$ and the running mass of the bottom quark 
$m_b(m_b)^{\overline   MS}$, all in the $\overline{MS}$ renormalisation scheme
The recent Tevatron top mass $m_t$ measurement \cite{mtmeas} is also employed. 
These SM inputs are shown in Table~\ref{tab:inp}. 
Experimental errors are so small on the mass of the $Z^0$ boson $M_Z$ 
and the muon decay constant $G_\mu$ that we 
fix them to their central values of 91.1876 GeV and $1.16637 \times 10^{-5}$
GeV$^{-2}$ respectively. 
The Standard Model (SM) input parameters are allowed to vary within 4$\sigma$
of their central values but a $\chi^2$ penalty 
\begin{equation}
\chi_i^2 = \frac{(c_i - p_i({\bf m}))^2}{\sigma_i^2} \label{chisq}
\end{equation}
is applied for observable $i$.  $c_i$ denotes the central value of the
experimental measurement, $p_i({\bf m})$ represents the value ``{\bf
p}redicted'' at any stage of the MCMC sampling given knowledge of the
model ${\bf m}$ presumed to be ``true'' at that point.  Finally
$\sigma_i$ is the standard error of the measurement.  Equivalently,
expressing this in the language of likelihoods, we are assuming that
each of these measurements have Gaussian errors,\footnote{The
experimental constraints on BR($B_s \rightarrow \mu^+ \mu^-$) and the
LEP constraints on the Higgs mass, each described later, are not
Gaussian constraints and must therefore be treated differently.
Nothing prevents us from continuing to parametrise their likelihood
distributions in the same way, however, but it should be realised that
a consequence of this is that the ``$\chi$-squared penalty'' (i.e. $-2
\log{{\mathcal L}_i}$) will not be parabolic, as ${\mathcal L}_i$ is
not a Gaussian distribution in these cases.} and that the likelihood
distribution ${\mathcal L}_i \equiv p(c_i | {\bf m})$ for any one
measurement may be written in the following way:
\begin{equation}
{\mathcal L}_i \equiv p(c_i | {\bf m}) =
\frac 1 {\sqrt{2 \pi \sigma_i^2}}
\exp \left[ {-\chi_i^2 /2} \right].
\end{equation}
The normalisation constant $\sqrt{2 \pi \sigma_i^2}$ may be ignored
in subsequent calculations as the absolute value of ${\mathcal L}_i$ will
never be needed.  It will only be necessary to know the {\em ratios}
\/of values of ${\mathcal L}_i$ at neighbouring points in the MCMC
chain, or between chains in which the neglected constants are identical.


In order to calculate predictions for observables from the inputs in
Table~\ref{tab:inp}, we use {\tt
  SOFTSUSY2.0.7}~\cite{softsusy} to first calculate the MSSM spectrum.
\TABULAR{|cc|cc|cc|cc|}
{\hline
$m_{\chi_1^0}$ & 37  & $m_{\chi^\pm_1}$ & 67.7 & 
$m_{\tilde g}$ & 195 &
$m_{{\tilde \tau}_1}$ & 76 \\ $m_{{\tilde l}_R}$ & 88 & $m_{{\tilde t}_1}$ &
  86.4 &
$m_{{\tilde b}_1}$ & 91 & $m_{{\tilde q}_R}$ & 250 \\
$m_{{\tilde \nu}_{e,\mu}}$ & 43.1 & & & & & & \\
\hline}{Lower bounds applied to sparticle mass predictions (in GeV). \label{tab:bds}}
We apply the bounds in
Table~\ref{tab:bds} in order to take into account 95$\%$ limits coming from
negative sparticle searches~\cite{PDG}. Any point transgressing these bounds 
is given a zero likelihood density (or, equivalently, an infinite $\chi^2$).
Also, we set a zero likelihood for any inconsistent point, e.g.\ one which does
not break electroweak symmetry correctly, or a point that contains tachyonic
sparticles. 
For points that are not ruled out,
we then link the MSSM spectrum via the SUSY Les Houches Accord~\cite{slha} to
  {\tt 
  micrOMEGAs1.3.6}~\cite{micromegas}, which then calculates $\Omega_{DM} h^2$, the
branching ratios $BR(b \rightarrow s \gamma)$ and $BR(B_s \rightarrow \mu^+
\mu^-)$ and $(g-2)_\mu$.  

The measured value of the anomalous magnetic moment $(g-2)_\mu$ is in conflict
with the SM predicted value by~\cite{PDG}
\begin{equation}
\delta a_\mu \equiv \delta \frac{(g-2)_\mu}{2} = (22 \pm 10) \times 10^{-10}.
\end{equation}
This excess may be explained by a supersymmetric contribution,
the sign of which is identical to the sign of the superpotential $\mu$
parameter~\cite{susycont}. After obtaining the one-loop MSSM value of
$(g-2)_\mu$ from {\tt micrOMEGAs}, we add the following dominant 2-loop
corrections~\cite{2loop,private}: the logarithmic piece of the 2-loop QED
contribution, two-loop stop-higgs and chargino-stop/sbottom contributions.

\EPSFIGURE{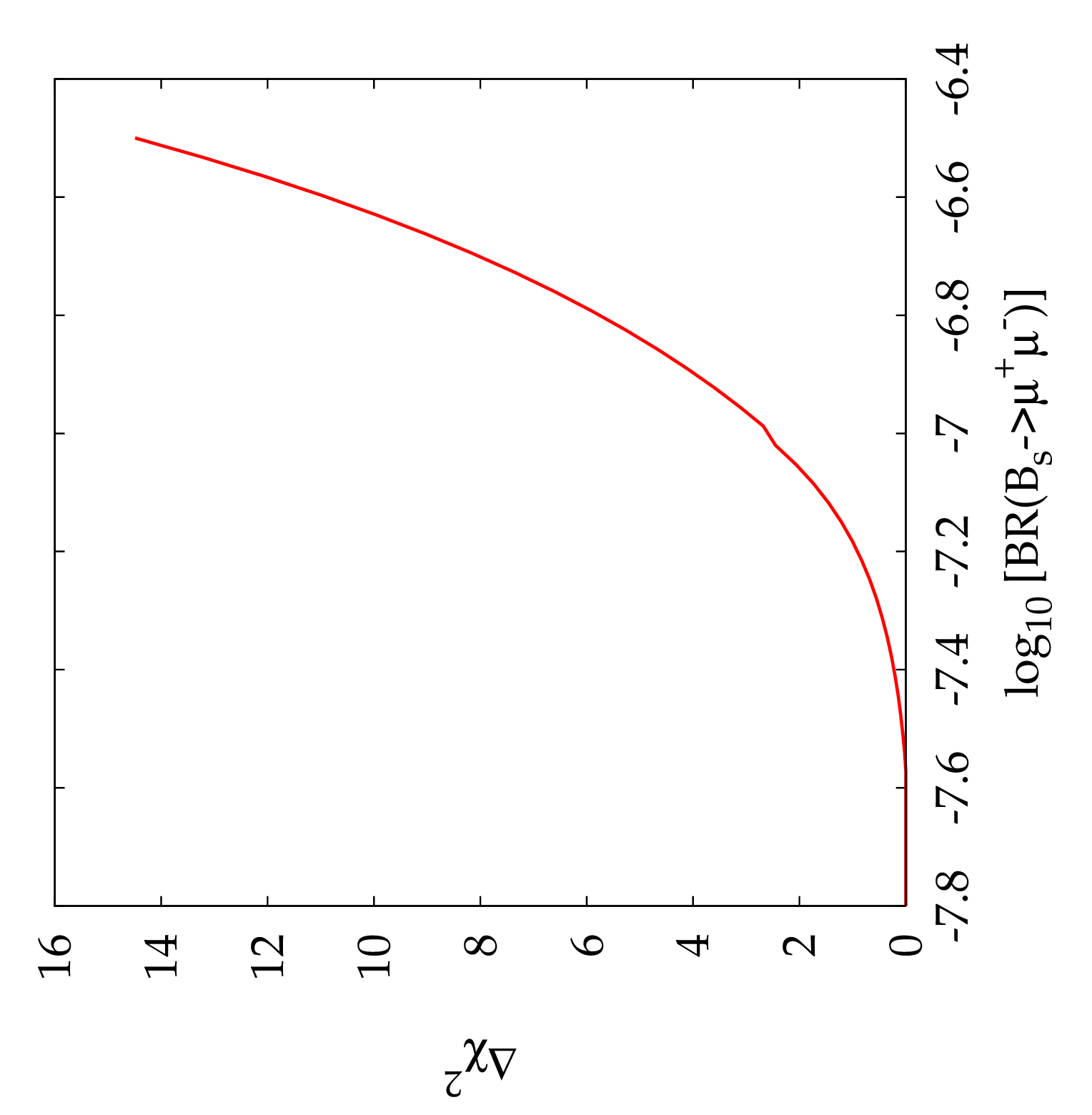,height=8cm,angle=270}{$\chi^2$ penalty on
  BR($B_s \rightarrow \mu^+ \mu^-$) from
  \cite{private2}\label{fig:bsmumu}} The Tevatron has recently been
  instrumental in bounding the branching ratio of the rare decay
  channel $B_s \rightarrow \mu^+ \mu^-$~\cite{bsmumu}.  Such bounds
  help constrain the mSUGRA parameter space~\cite{Ellis:2005sc}.  We
  apply a $\chi^2$ penalty on the value predicted by {\tt
  micrOMEGAs1.3.6} derived from CDF Tevatron Run II
  data~\cite{private2}. The resulting penalty is shown in
  Fig.~\ref{fig:bsmumu}.

Recently, it has been claimed that light sparticles are preferred by
the two weak observables $\sin^2 \theta_w^l$ and $M_W$
\cite{Ellis:2004tc,Ellis:2006ix}.  In Ref.~\cite{deAustri:2006pe},
$M_W$ and $\sin^2 \theta_w^l$ were used at one loop order to help
constrain mSUGRA in a global MCMC fit. The preference for such light
SUSY was not particularly evident in the global fits.  We examine the
mSUGRA predictions for $M_W$ and $\sin^2 \theta_w^l(\mbox{eff})$ in
Figs.~\ref{fig:mw} and~\ref{fig:st} for $A_0=0$, $\tan \beta=10$,
$\mu>0$, equal $m_0$ and $M_{1/2}$ and central experimental values for
the other inputs.  For the ``SOFTSUSY'' lines, the default {\tt
SOFTSUSY} calculation is used.  This contains the full SOFTSUSY MSSM
contributions to the leptonic mixing angle $\sin^2 \theta_w^l$ and
$M_W$. It also contains the dominant 2-loop Standard Model
contributions to $M_W$.  For the lines marked ``2-loop'', the SUSY Les
Houches Accord is used to communicate with a currently private code
that calculates the W-boson mass $M_W$~\cite{Heinemeyer:2006px}, and
the effective leptonic mixing angle variable $\sin^2\theta^l_w$,
calculated to two loops in the dominant MSSM parameters. We use the
most general MSSM result for the full one-loop contributions.  Besides
all known corrections due to SUSY particles, the full SM contributions
are also included in the predictions for $M_W$ and $\sin^2\theta^l_w$,
leading to the currently most accurate predictions within the MSSM\@.
The ``SM'' lines show the SM limit, where all corrections involving
sparticles are dropped, i.e.\ the state-of-the-art SM results
\cite{MWSM,Awramik:2004ge} with $M_{H^{\mathrm{SM}}}=M_h$.  They vary
slightly with $m_0=M_{1/2}$ because the varying mSUGRA parameters
produce different values of the Higgs boson mass $m_h$.  The
horizontal lines on the figures show the current $1\sigma$
experimental limits~\cite{mw,sinth}
\begin{equation}
M_W = 80.392\pm0.031\mbox{~GeV}, 
\qquad \sin^2 \theta_w^l = 0.23153 \pm 0.00020,
\end{equation}
where we have added experimental and theoretical errors in
quadrature. 

\DOUBLEFIGURE{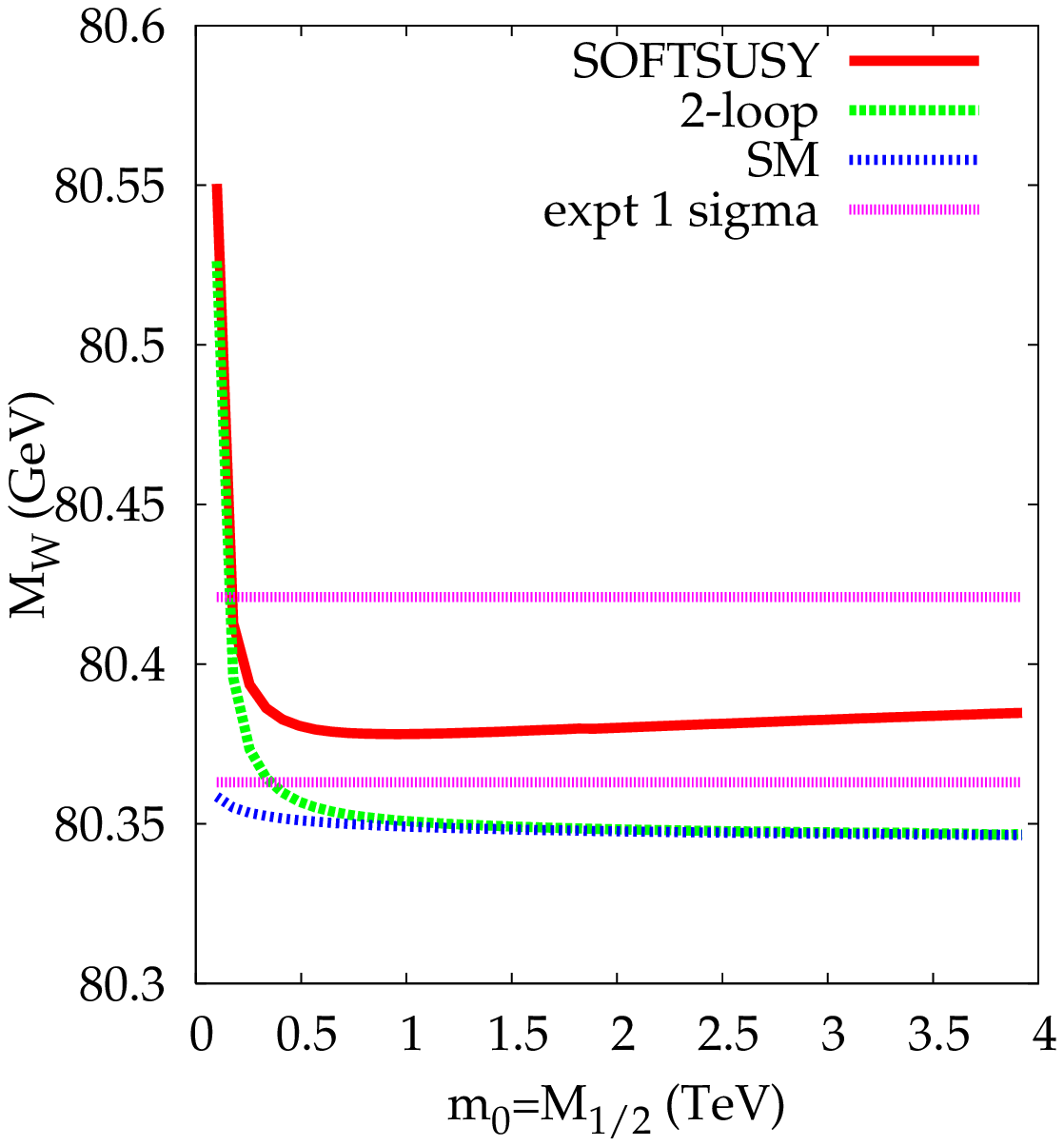,height=8cm}{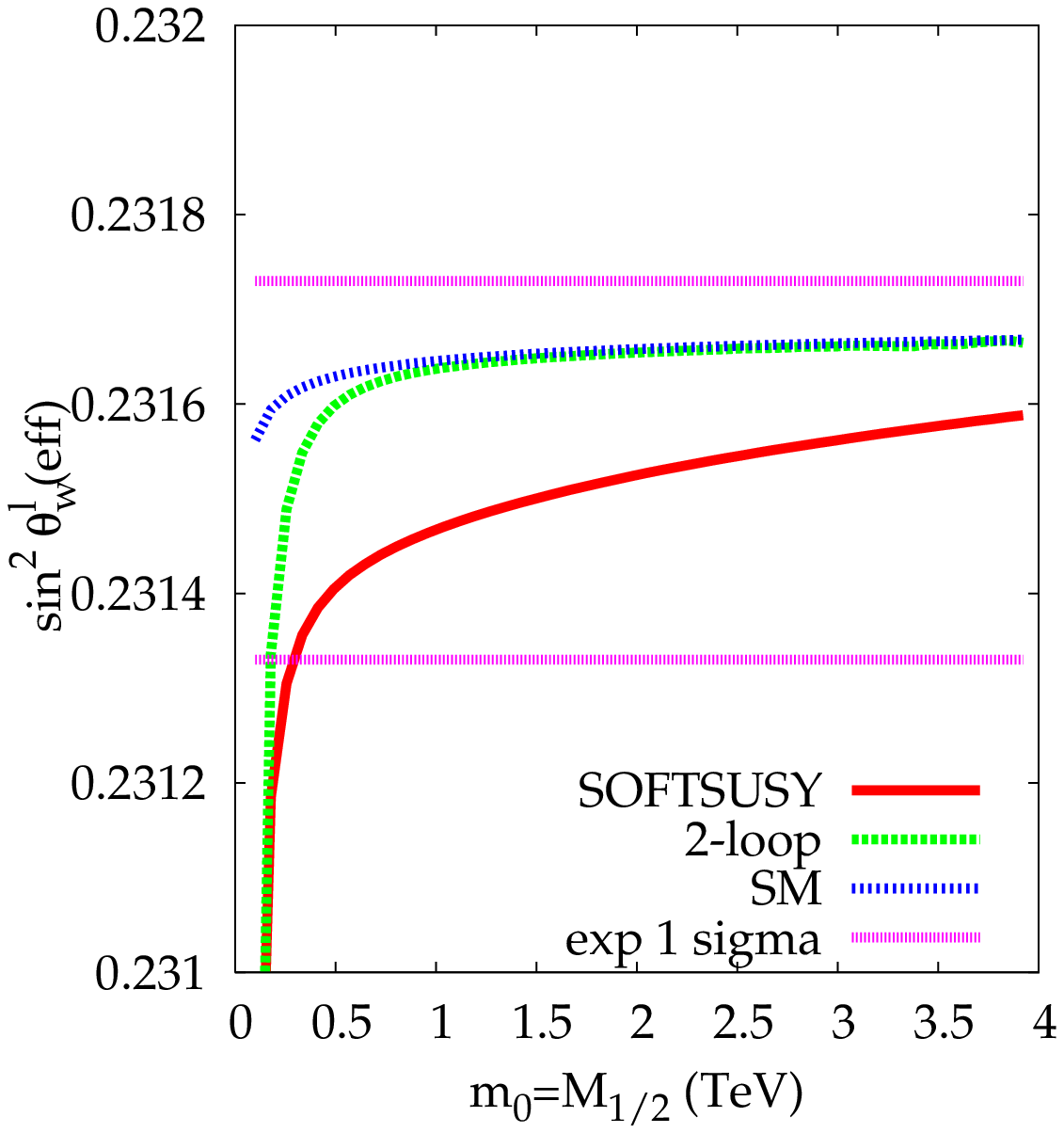,height=8cm}{Various approximations to $M_W$
  in mSUGRA \label{fig:mw}}{Various
  approximations to $\sin^2 \theta_w$ in mSUGRA. \label{fig:st}}
The theoretical errors in the predicted $M_W$ and $\sin^2
\theta_w^l$  are estimated to be $10$ MeV and 12$\times 10^{-5}$
respectively~\cite{Ellis:2003si}. We use these uncertainties for the
purposes of comparison, although they have been slightly reduced recently by
  the addition of additional two-loop corrections taken into account in the
  present analysis~\cite{Heinemeyer:2006px,drMSSMal2B}. 
We see from the SOFTSUSY line in the figure that
the  prediction of $M_W$ does not have a strong preference for light SUSY,
since the model is within the 1$\sigma$ errors up until $m_0=M_{1/2}=4$
TeV. In actual fact, only very light SUSY masses are disfavoured by the
``SOFTSUSY'' line, leading to predictions above the $1\sigma$-range.   
The situation is similar for the $\sin^2\theta^l_w$ ``SOFTSUSY'' line, where
again only very light SUSY masses lead to predictions outside the
$1\sigma$-range. %
However, using the best available predictions, corresponding to the ``2-loop''
lines, a 
preference for light $m_0=M_{1/2}$ can be seen in the prediction for
$M_W$. 
The SM curve which lies just below the $1\sigma$-interval is approached
from above in the decoupling limit, furthermore indicating a slight preference
of the MSSM over the SM\@. 
The preference for light SUSY is not as striking for $\sin^2\theta_w^l$. Here
most of the $m_0=M_{1/2}$ values are doing equally well, which is mainly due
to the fact 
that the SM prediction for $\sin^2\theta_w^l$ is already well within the
$1\sigma$-range.
With the behaviour of the ``one-loop'' curve and the best available result
being qualitatively different, it is
desirable to use the more accurate result for $M_W$ and
$\sin^2\theta_w^l$  when calculating 
the $\chi^2$ contributions of $M_W$ and $\sin^2 \theta_w^l$  using
Eq.\ref{chisq}. Although Ref.~\cite{deAustri:2006pe} only used the one-loop
predictions, the theoretical errors were correspondingly enlarged in order to
take the larger uncertainty from higher order terms into account.

  \EPSFIGURE{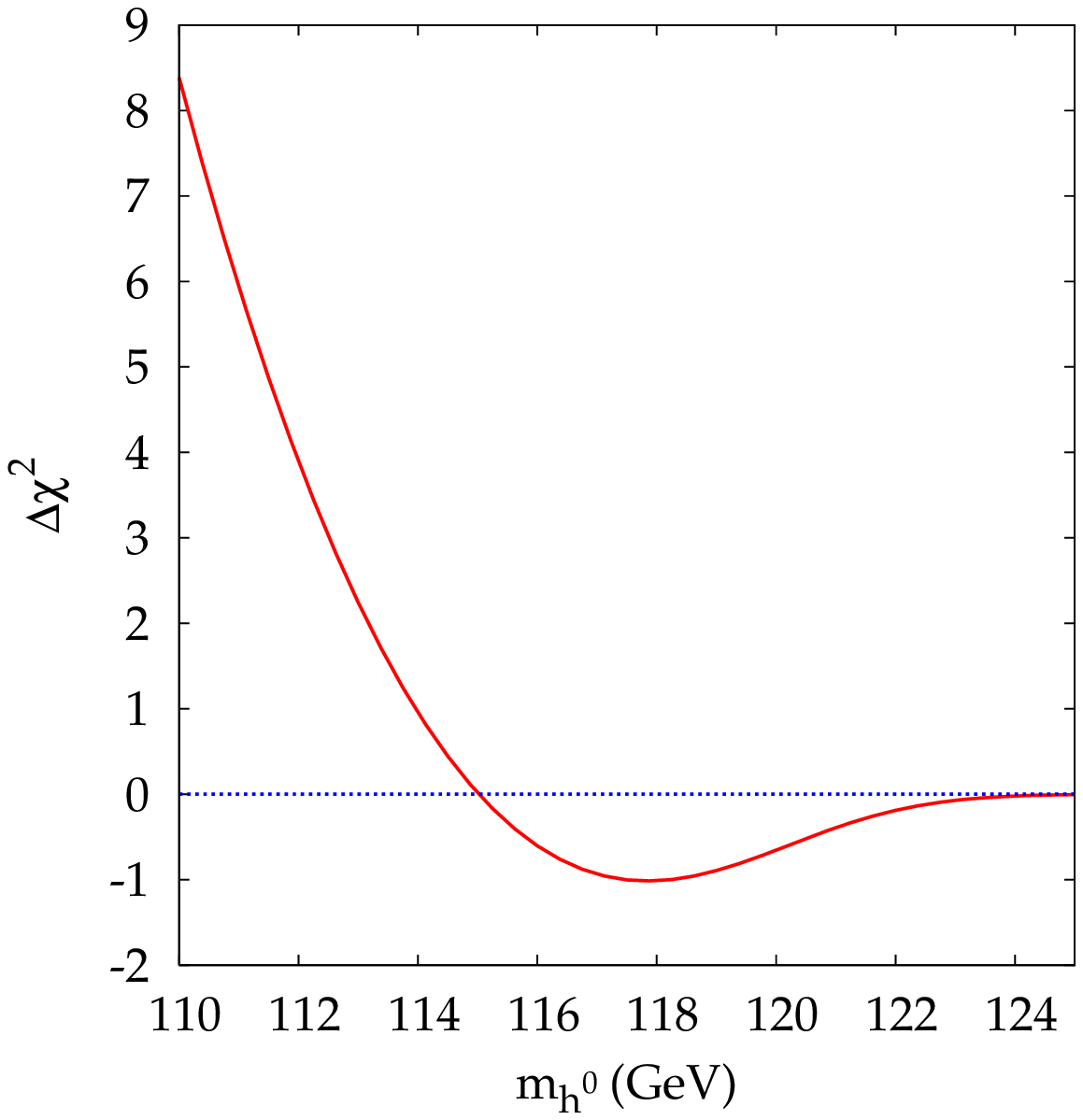,height=7cm}{LEP2 higgs $\chi^2$ penalty paid 
\label{fig:mhpen}}
LEP2 constraints on the lightest CP-even higgs mass are included as a further
likelihood penalty following a parameterisation of LEP2 data in the SM
limit~\cite{lep2higgs}. For the LEP2 constraints,
the SM limit is a good approximation for mSUGRA, since 
sparticle mass limits imply that we must be near the decoupling r\'{e}gime of
the MSSM~\cite{decouple}. We estimate that the {\tt SOFTSUSY2.0.7}
determination of $m_h$ has a 2 GeV theoretical
error in mSUGRA, although it may be somewhat larger in the general
MSSM~\cite{Allanach:2004rh}.  
We therefore smear the parameterised LEP2 Higgs likelihood density ${\mathcal
  L}_{LEP2}(m_h)$
with a Gaussian distribution of width $\sigma_h=2$ GeV:
\begin{equation}
{\mathcal L}_h(m_h) = \int_{m_h-4 \sigma_h}^{m_h+4 \sigma_h} dx \frac{1}{\sqrt{2 \pi}
  \sigma_h} e^{\frac{-(m_h-x)^2}{2 
  \sigma_h^2}} {\mathcal L}_{LEP2}(x).
\label{Higgserror}
\end{equation}
The result of this procedure
  leads to the effective $\Delta \chi^2=-2 \ln {\mathcal L}_h$ penalty shown in
  Fig.~\ref{fig:mhpen}. The slight excess of candidate Higgs events
  over the background prediction at LEP2 can be seen by a negative
  $\Delta\chi^2$ penalty in the figure for $m_h \sim 116-121$ GeV.

The rare bottom quark branching ratio is
$BR(b \rightarrow s \gamma)$ is constrained to be~\cite{hfg}
\begin{equation}
BR(b \rightarrow s \gamma)=  (3.55\pm0.38) \times 10^{-4}, \label{bsg}
\end{equation}
obtained by adding the experimental error with the estimated theory
error~\cite{gamb} of $0.3 \times 10^{-4}$ in quadrature.
Very recent estimates~\cite{Misiak:2006zs,Andersen:2006hr}
of $BR(b \rightarrow s \gamma)$ are compatible with Eq.~\ref{bsg} at the
1$\sigma$ level, although
the error has decreased.
Our prediction of $BR(b \rightarrow s \gamma)$ is substituted for $p_i$ in
Eq.~\ref{chisq} in order to calculate $\chi_{BR(b \rightarrow s \gamma)}^2$.

We use the WMAP3~\cite{wmap} power law $\Lambda$-CDM fitted value of 
\begin{equation}
\Omega \equiv \Omega_{DM} h^2 = 0.104^{+0.0073}_{-0.0128} \label{omega}
\end{equation}
for the dark matter relic density of the universe. We initially assume that the
neutralinos are stable and that they constitute the whole of the dark matter
relic density. 
Eq.~\ref{chisq} is used to calculate $\chi^2_{\Omega}$, with
$\sigma_{\Omega}=0.0073$ for a prediction lower than the central
experimental value and 0.0128 otherwise.

Having described the calculation of the likelihood associated
with each individual measurement, we are now in a position to define
the likelihood of the set of all measurements or observables, taken
together.  We are required to calculate the joint (total) likelihood
$\mathcal L$ of all the measurements given the truth, {\em i.e.}\ in the
notation of Eq~\ref{chisq} we want to know:
\begin{eqnarray}
{\mathcal L} &=& p( \mbox{measurements} | \mbox{true model}) \\
&=& p(c_1, c_2, \ldots | {\bf m}) \\
&=& p(c_1 | {\bf m}) \cdot p(c_2 | c_1 , {\bf m}) \cdot p(c_3 |
c_1, c_2, {\bf m}) \cdot \ldots \\
&=& {\mathcal L}_1 \cdot p(c_2 | c_1 , {\bf m}) \cdot p(c_3 |
c_1, c_2, {\bf m}) \cdot \ldots
\end{eqnarray}
which is not in general equal to
\begin{equation}
{\mathcal L}_1 \cdot
{\mathcal L}_2 \cdot
{\mathcal L}_3 \cdot
\ldots
\end{equation}
unless we can be confident that for each measurement
\begin{equation}
p(c_3 | c_1, c_2, {\bf m}) \approx p(c_3 | {\bf m}) \mbox{\qquad{\em
etc}}.\label{eq:whatwewantapprox}
\end{equation}
Fortunately we can be confident that Eq~\ref{eq:whatwewantapprox} {\em
does}\/ hold in our situation, as we have deliberately constructed ${\bf
m}$ to be broad enough such that all fundamental parameters which
might reasonably be expected to correlate any two of the measurements
are included within it\footnote{If the design of ${\bf m}$ were not
broad enough, then ${\bf m}$ would have to be extended.  For example:
were it the case that the up quark mass $m_u$ was expected to 
significantly correlate two
or more of the observables, then for Eq~\ref{eq:whatwewantapprox} to
continue to hold, $m_u$ would have to be added to ${\bf m}$ enlarging
its dimension by one.}.  We are therefore at liberty to write:
\begin{eqnarray}
{\mathcal L} &=& {\mathcal L}_1 \cdot
{\mathcal L}_2 \cdot
{\mathcal L}_3 \cdot
\ldots \\ 
&=& e^{-\sum_i \chi_i^2/2}
\end{eqnarray}
for the total likelihood, and we can be confident that this product
takes into account the expected correlations between all the
observables contained, due to the nature of the space $\left\{{\bf m}
\right\}$ of models considered.

\section{Dark Side Fits \label{sec:dark}}

\FIGURE{\twographst{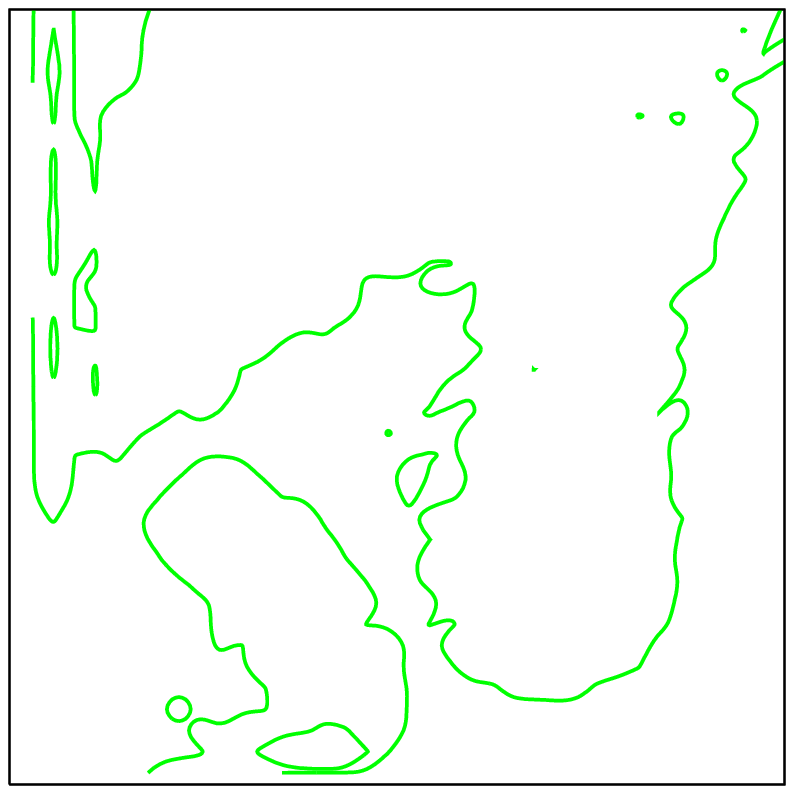}{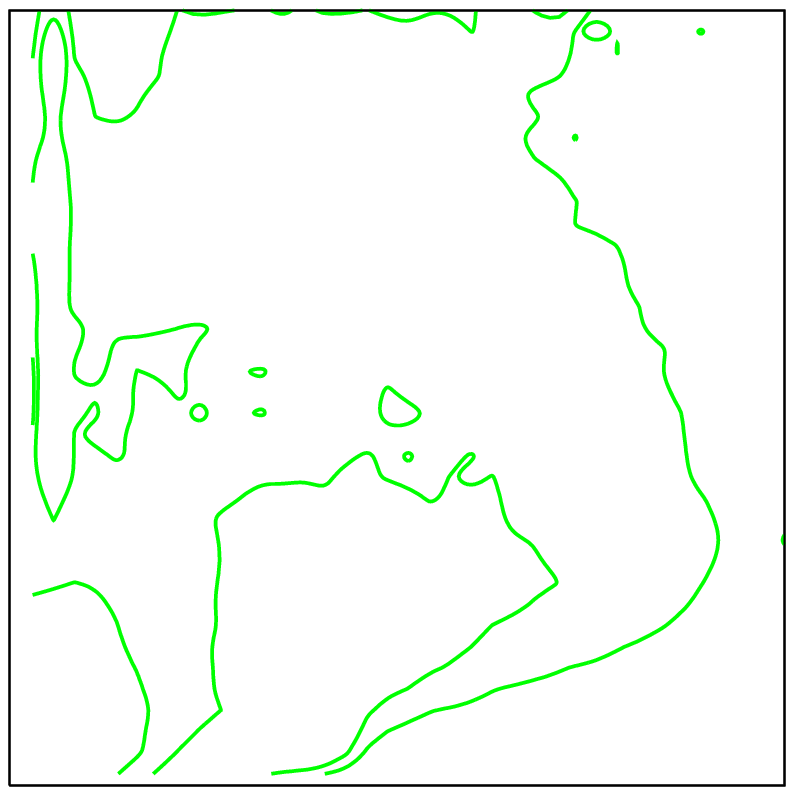}
\caption{mSUGRA Fits for (a) $\mu<0$ (b) $\mu>0$ marginalised to the
  $m_0$-$M_{1/2}$ plane. The posterior probability is indicated by the
  bar on the right hand side. The inner (outer) contours show the
  $68\%$ $(95\%)$ confidence region respectively.\label{fig:m0m12}} }
  We now compare and contrast the dark side fits to those with
  $\mu>0$.  In Figs.~\ref{fig:m0m12}(a) and~\ref{fig:m0m12}(b), we
  show the posterior probabilities $P$ marginalised\footnote{For readers
  unfamiliar with the term: marginalisation
  means ``integrated over the unseen dimensions of parameter space'' in this
  context.} to the 
  $m_0-M_{1/2}$ plane for both signs of $\mu$.  As with all
  2-dimensional marginalised plots in this paper, we bin the plane into
  75$\times$75 2-dimensional bins. The colour bar on the right hand side
  of the figures shows the posterior probability $P$ of each bin divided
  by the maximum posterior probability of any bin in the plot. In
  Fig.~\ref{fig:m0m12}(a), the only 68$\%$ contour\footnote{Note that the
    confidence regions in Figs.~\ref{fig:m0m12} (and those in later plots)
    should strictly be referred to as ``Bayesian credible intervals''
    (each region contains a fixed amount of the posterior probability) to
    distinguish them from the related concept from Frequentist Statistics
    called a ``confidence interval''. Usage of the term ``credible
    interval'' is not common in High Energy Physics, however, and we will
stick to ``confidence region''.} is at the lowest
  values of $m_0$: all other contours are 95$\%$ confidence region contours
  due to   the lowish likelihood densities. The brokenness of the contours is a
  result of statistical fluctuations in the results. Although
  these are visible, there is clearly reliable information 
  their trajectories in the plot.  The $\mu<0$ plot displays
  two isolated local maxima,
  whereas the 95$\%$ confidence region of $\mu>0$ is continuously
  connected. We discuss the relative normalisation of the two $\mu<0$ maxima
  in appendix~\ref{sec:conv}.
  In the 95$\%$ confidence region of Fig.~\ref{fig:m0m12}a closest to the
  origin,   the  relic 
  density is dominantly depleted 
  by either stop co-annihilation~\cite{Boehm:9911,arnie,Ellis:2001nx} ${\tilde
  t} \chi_1^0   \rightarrow t g$ or  
  stau co-annihilation~\cite{Griest:1990kh} ${\tilde \tau}_1 \chi_1^0 \rightarrow \tau \gamma$.
On the other hand, the 95$\%$ region at higher $m_0-M_{1/2}$ 
consists of resonant Higgs annihilation
regions~\cite{Drees:1992am,Arnowitt:1993mg,Djouadi:2005dz}, where
$\chi_1^0\chi_1^0 \rightarrow h,A^0 \rightarrow b {\bar b}/\tau^+
\tau^-$ and the focus point region where the LSP has a significant
higgsino component and $\chi_1^0\chi_1^0 \rightarrow ZZ, WW, t
\bar{t}$~\cite{Feng:1999mn,Feng:1999zg,Feng:2000gh}.
There was no significant
stop co-annihilation~\cite{Boehm:9911,arnie,Ellis:2001nx} region for
$\mu>0$.  

\FIGURE{\fourgraphst{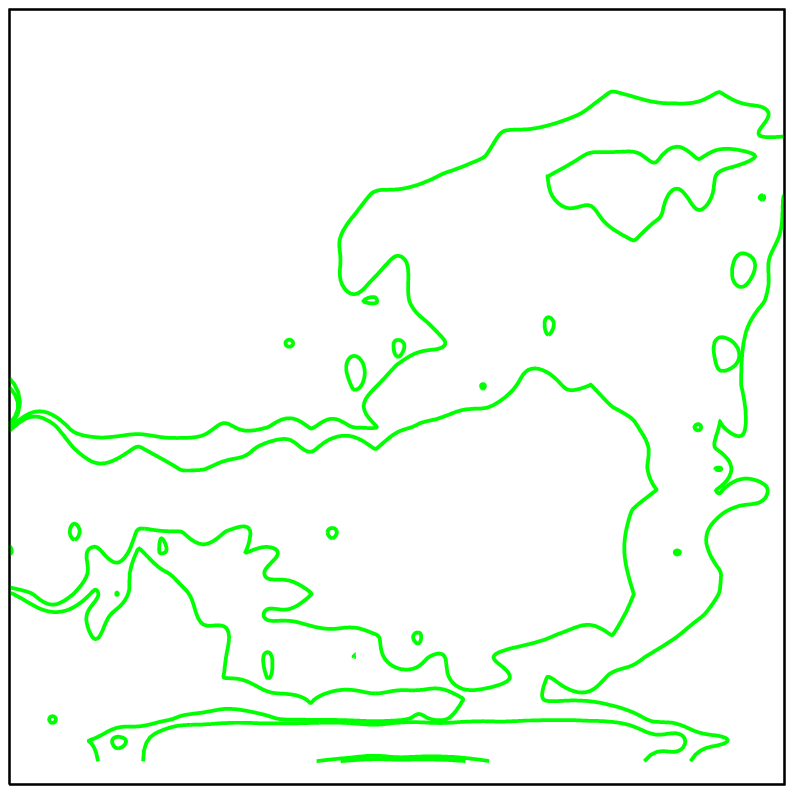}{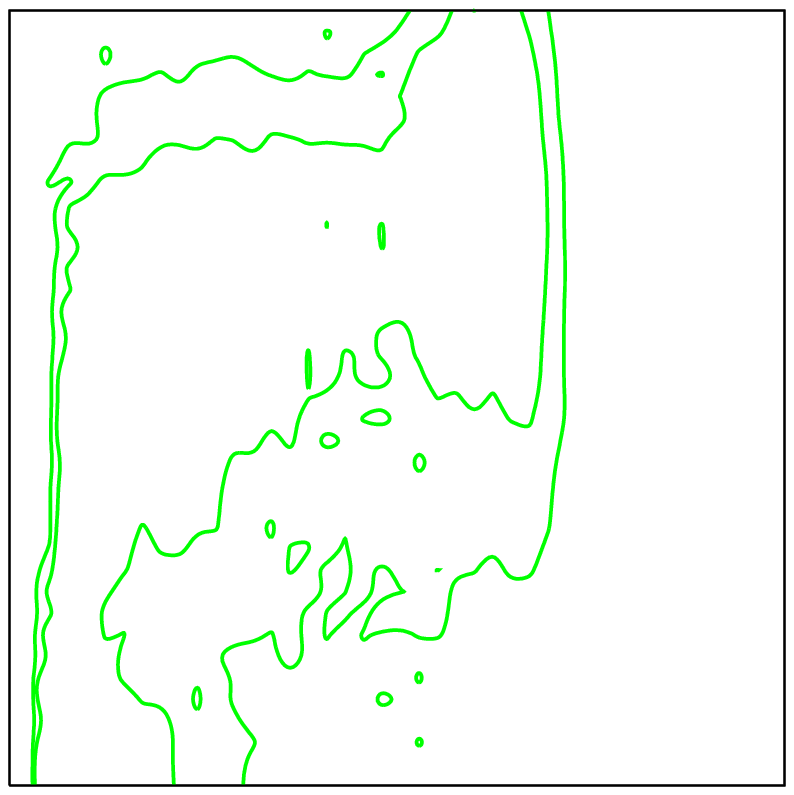}{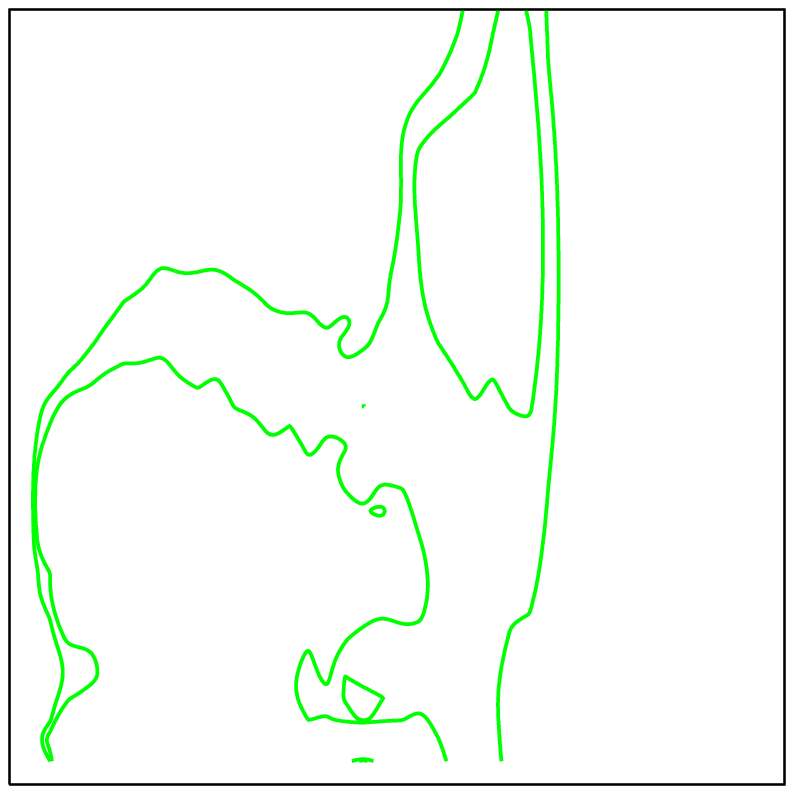}{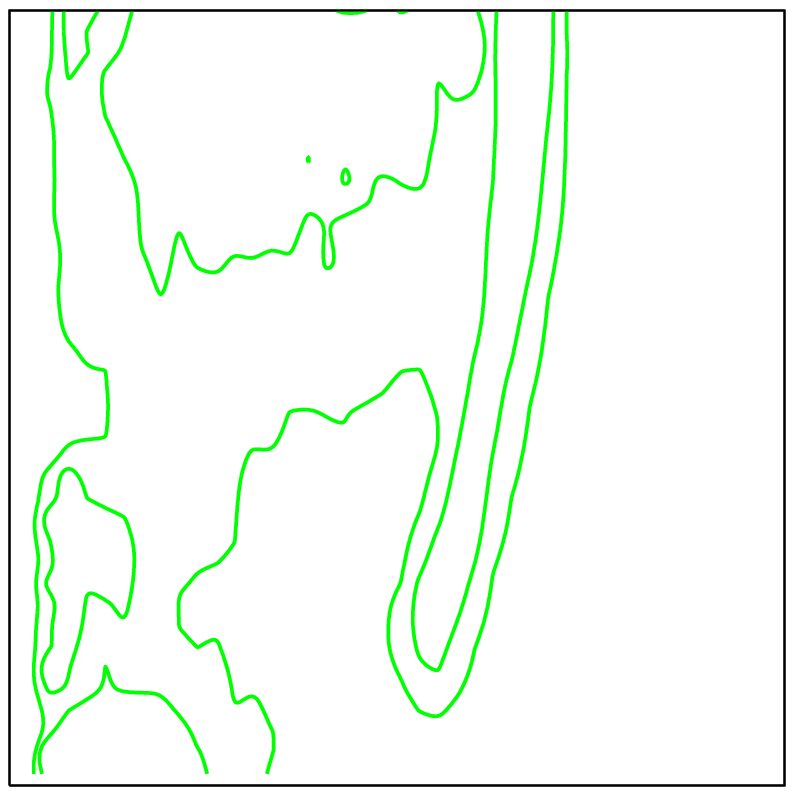}
\caption{Constraints from global fits with $\mu<0$ mSUGRA marginalised to
  2-dimensional parameter planes.
We have assumed a flat prior. The posterior probability is indicated by the bar
  on the right hand side. The inner (outer) contours show the
  boundary of a $68\%$
  $(95\%)$ confidence region.
\label{fig:mulz}}
} We now include some other marginalisations on the other
2-dimensional parameter planes for $\mu<0$ mSUGRA with a flat
prior. They are displayed in
Figs.~\ref{fig:mulz}(a)-(d). Figs.~\ref{fig:mulz}(a) and~\ref{fig:mulz}(b)
show that the probability density for the 2-chain 
co-annihilation sample is not separated from the other sample in either
the $M_{1/2}-A_0$ plane or the $\tan \beta$-$A_0$ plane (the almost
disconnected region at the bottom of Fig.~\ref{fig:mulz}(a) consists
of the light $h^0$-pole region). There is only a modest separation in
the $M_{1/2}$-$\tan \beta$ plane: the 68$\%$ contours do not connect
the two regions, whereas the 95$\%$ contours
do. Fig.~\ref{fig:mulz}(d) shows that the connection between the two
samples in the $m_0-\tan \beta$ plane is marginal. The $m_0-A_0$
marginalisation was useful for investigating the physics behind the two
isolated probability maxima, and is displayed in appendix~\ref{sec:conv}.

\FIGURE{\fourgraphs{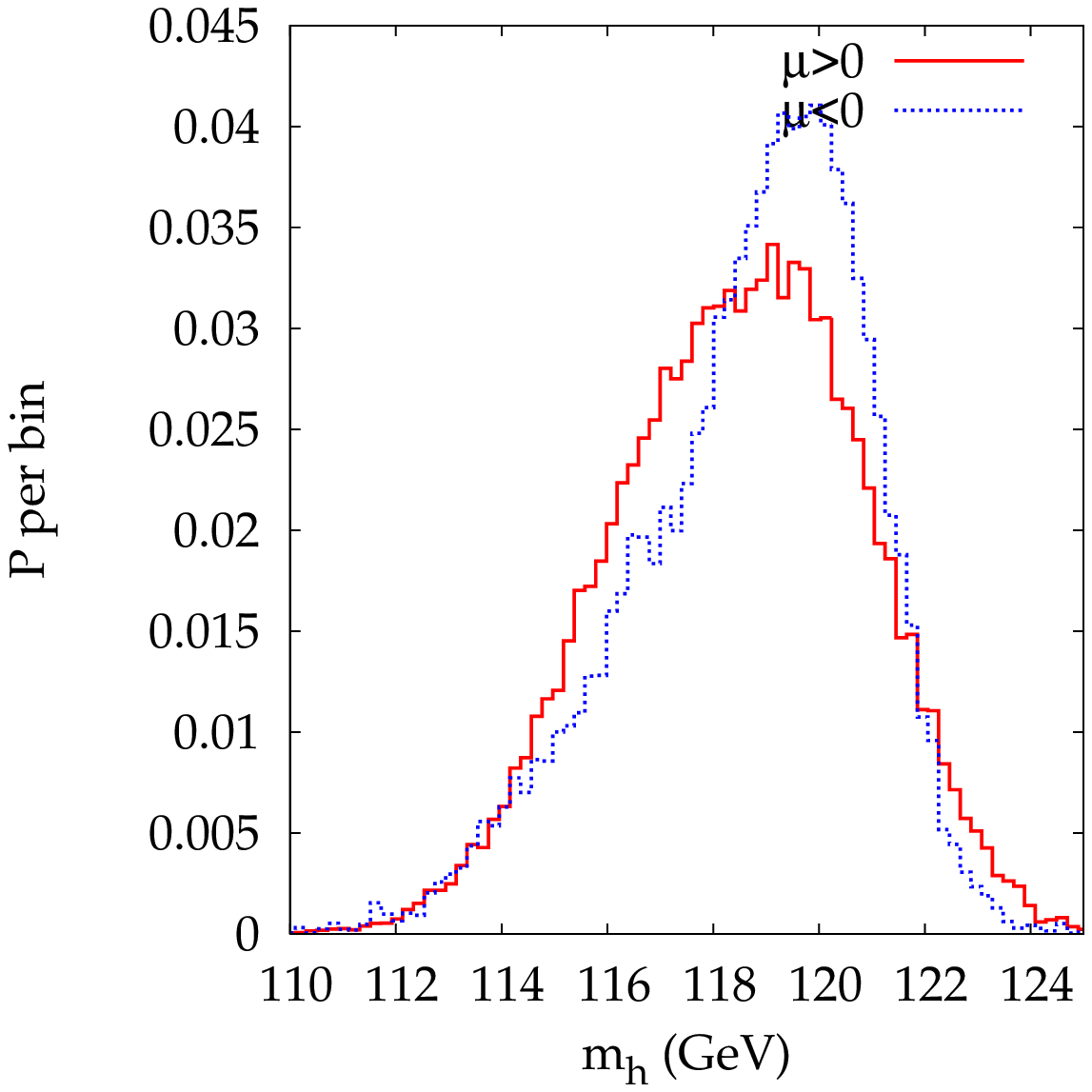}{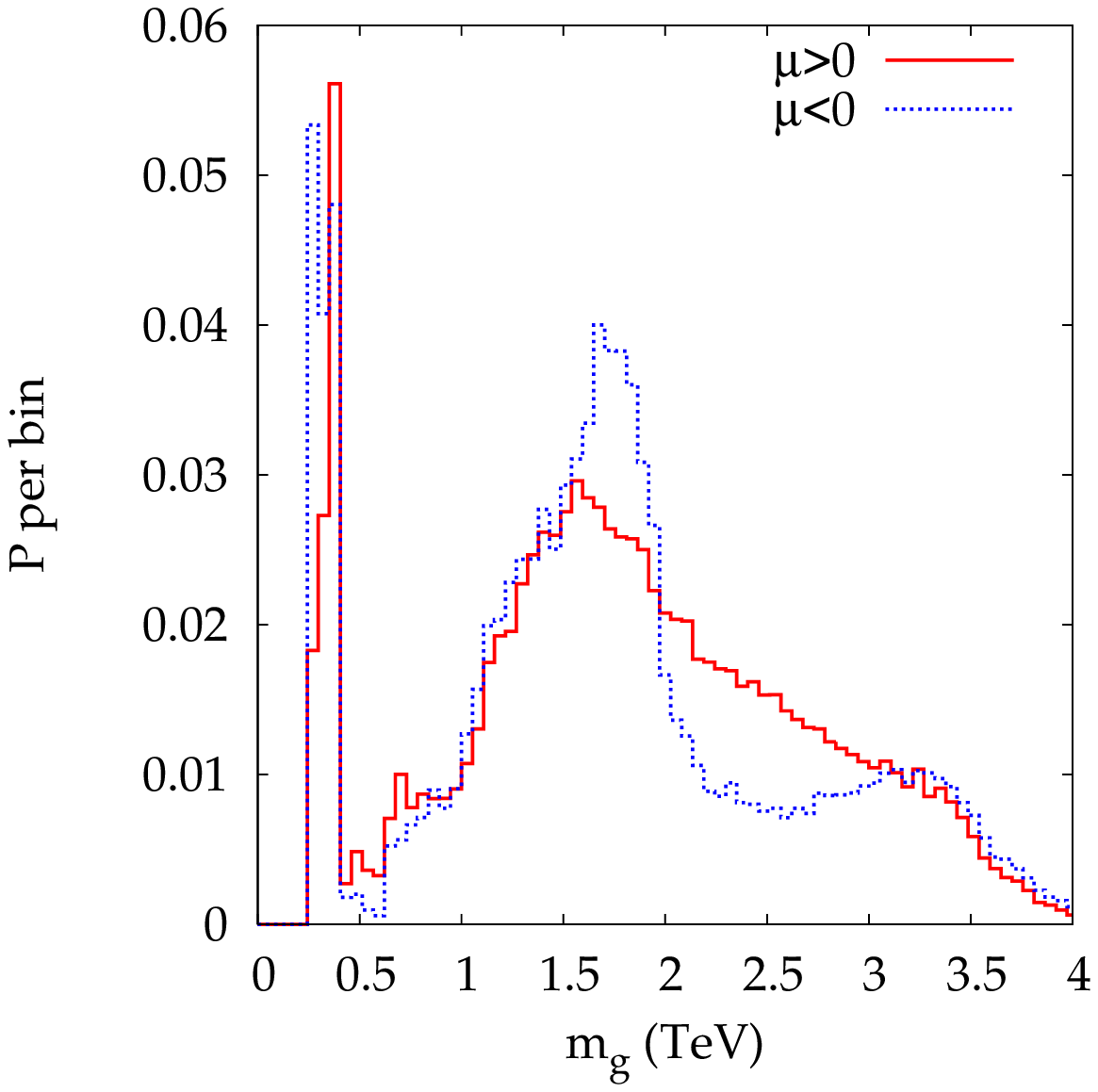}{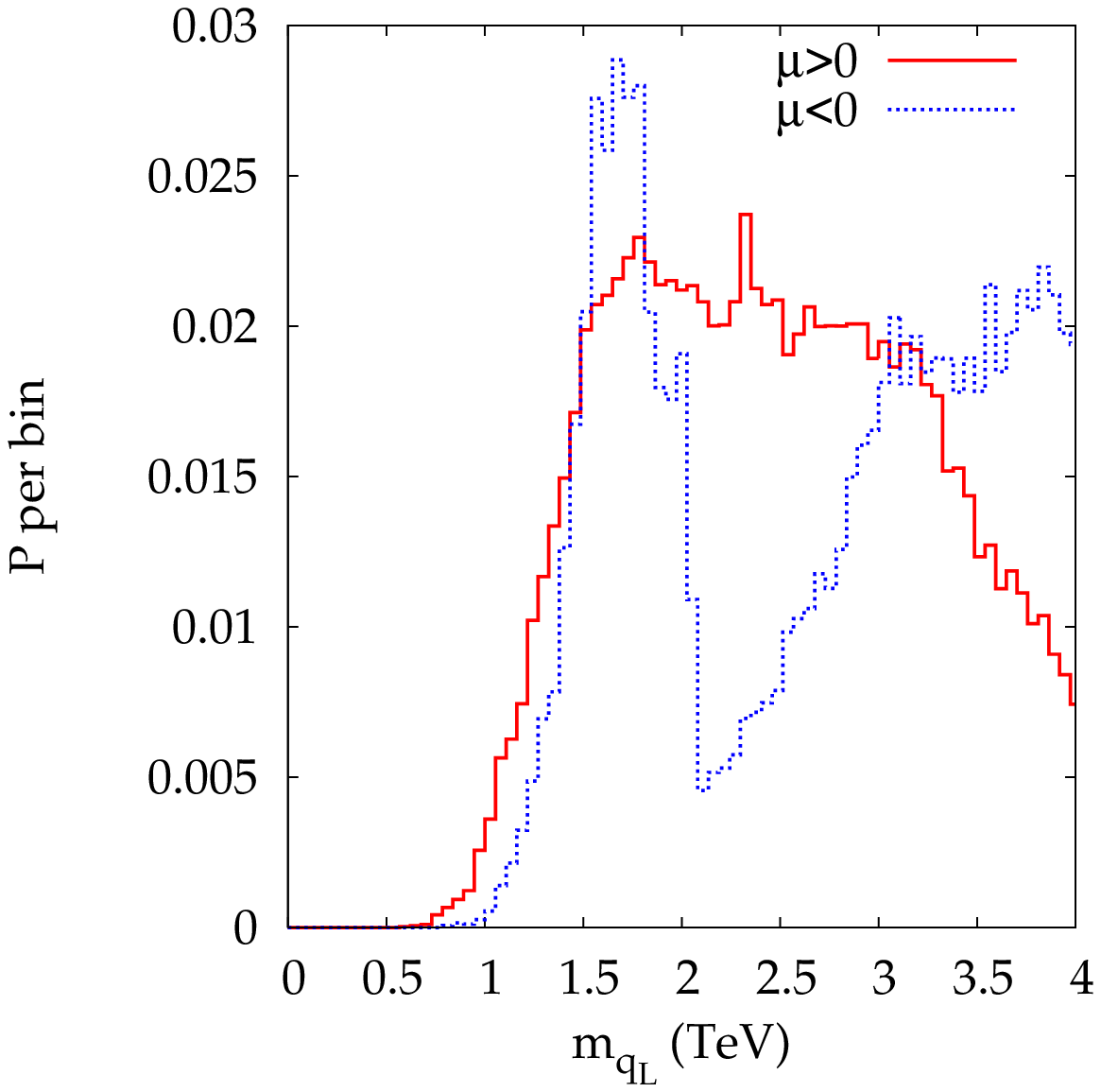}{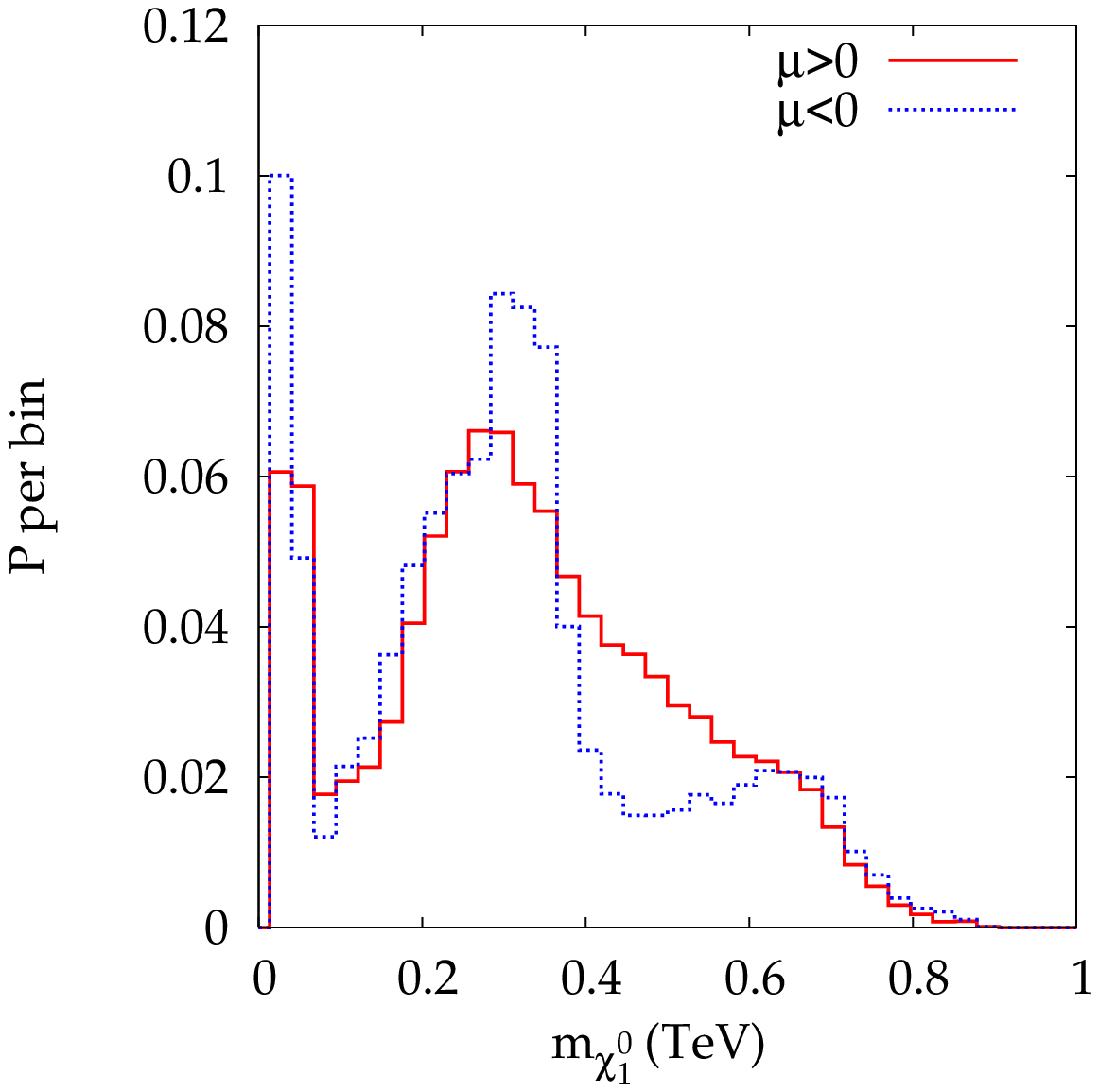}
\caption{Probability distributions in mass of (a) the lightest CP even higgs,
 (b) gluino, (c) the left-handed squark and (d) the neutralino. Flat priors
 have been assumed.
\label{fig:mass}}
}
The probability distributions of the masses of selected MSSM particles are
shown in Fig.~\ref{fig:mass} for
both signs of $\mu$ in mSUGRA\@. 
The $\mu<0$ sample has not been
normalised to the correct relative normalisation compared to $\mu>0$ in the
figure. The lightest CP-even higgs, the gluino and the lightest neutralino all
have mass distributions that are remarkably similar for either sign of $\mu$. 
However, in Fig.~\ref{fig:mass}(c), we see that the $\mu<0$ sample has a flat
plateau for higher squark masses, whereas the $\mu>0$ sample tails off
somewhat. High squark masses
will result in smaller total SUSY cross-sections at the LHC but the lighter
gluino should still provide enough events for discovery if $m_{\tilde g}<2$
TeV~\cite{Armstrong:1994it,CMS}.
Sharp peaks at low gluino and neutralino masses are due to the
$h^0$-resonance annihilation region~\cite{Allanach:2005kz}. The broader peak
of the $\mu<0$ curve in Fig.~\ref{fig:mass}(c) is due mostly to the
co-annihilation sample. The significant probability densities for large gluino
and squark masses are rather alarming, as then SUSY detection at the LHC would
require a longer running time.
Gauginos are not so sensitive to the range of the
prior in Table~\ref{tab:inp} but the sfermions are~\cite{deAustri:2006pe} and
a reduced range makes lighter sfermions more likely.
Also, naturalness priors~\cite{Allanach:2006jc} have a large impact, reducing
the likely sparticle masses. 

\subsection{Further investigations of the fits}

We examine the best-fit points from each sampling in Table~\ref{tab:bestfit}.
\TABULAR{|c|cc|c|cc|}{\hline
      & $\mu<0$ & $\mu>0$ & & $\mu<0$ & $\mu>0$  \\ \hline
$m_0$/GeV &  3610   & 156   & $\delta a_\mu/10^{-10}$ &-0.4 & 14.2 \\
$M_{1/2}$/GeV & 93 & 569  & $BR(b \rightarrow s \gamma)/10^{-4}$ &3.65 &3.41 \\
$A_0$/GeV &   -56   & 270  & $BR(B_s \rightarrow \mu^+ \mu^-)/10^{-9}$ &3.1 &3.7 \\
$\tan \beta$ &6.0 &24.1 & $\sin^2 \theta_w^l(\mbox{eff})$ & 0.23153&0.23152\\
$\Omega_{DM} h^2$ & 0.102 & 0.101 & $M_W$/GeV & 80.382& 80.368\\
$m_h$/GeV & 117.5 & 115.8 & $\chi^2$ & 4.5& 1.5\\
\hline
}{Best-fit points from the MCMC samplings for each sign of $\mu$. 
$\chi^2 \equiv
  \sum_i \chi_i^2$ and the Standard Model inputs are close to their
  experimental central values in each case. \label{tab:bestfit}}
We see from the table that, as expected, the $\mu>0$ sample has a better
best-fit point and a  correspondingly lower $\chi^2$, mainly due to the much
better fit to $\delta a_\mu$. In agreement with
Refs.~\cite{Ellis:2004tc,Ellis:2006ix}, the best-fit $\mu>0$ point is at light
SUSY masses. 
The $m_0$ and $M_{1/2}$ parameters are smaller
for the $\mu>0$ case than for $\mu<0$, corresponding to lighter sparticles and
the larger contribution to the anomalous magnetic moment of the muon. 
If we take the $\mu<0$ best-fit point and {\em flip}\/ the sign of $\mu$, we
find that the point does not break electroweak symmetry properly and is
excluded. The $\mu>0$ best-fit 
point has a total $\chi^2$ of 181 when the sign of $\mu$ is flipped, mostly
due to an increase in the predicted value of $\Omega_{DM} h^2$ to 0.195.

The region of smaller $m_0$, $M_{1/2}$ is more
  probable for $\mu>0$ than for $\mu<0$: this will lead to a relatively
  heavier $\mu<0$ 
  spectrum.  Our $\mu>0$ results are generally similar to previous analyses
  which did {\em not} \/include $M_W$ and $\sin^2 \theta_w^l$ as
  constraints in Ref.~\cite{Allanach:2005kz} and to those which included the
  one-loop {\tt SOFTSUSY} prediction for $M_W$, $\sin^2 \theta_w^l$ with
  enlarged theoretical errors~\cite{deAustri:2006pe}.  
  This seems in contradiction to the conclusions of
  Ellis {\em et al}~\cite{Ellis:2004tc,Ellis:2006ix}, where it is claimed that
  the electroweak   variables prefer light SUSY\@. Indeed, Fig.~\ref{fig:st}
  indicates that $M_W$, $\sin^2 \theta_w^l$ do mildly prefer light SUSY but
  our results show that this preference is washed out in the global fits. 
  Our results allow for heavier sparticles than Ellis {\em et al}, mainly
  because we have chosen to allow more relevant parameters to vary: 8 compared
  to 2 in their paper (one dimension is fixed by requiring the relic density
  prediction to be the central WMAP-constrained value). Their fits are for
  different discrete values of 
  fixed $\tan \beta$, but if it were allowed to vary, we believe that the
  confidence regions there would be enlarged. In the present paper, we also
  obtain additional smearing from allowing $m_t$, $\alpha_s$, $\alpha$
  and $m_b$ to vary.
  \FIGURE{\twographs{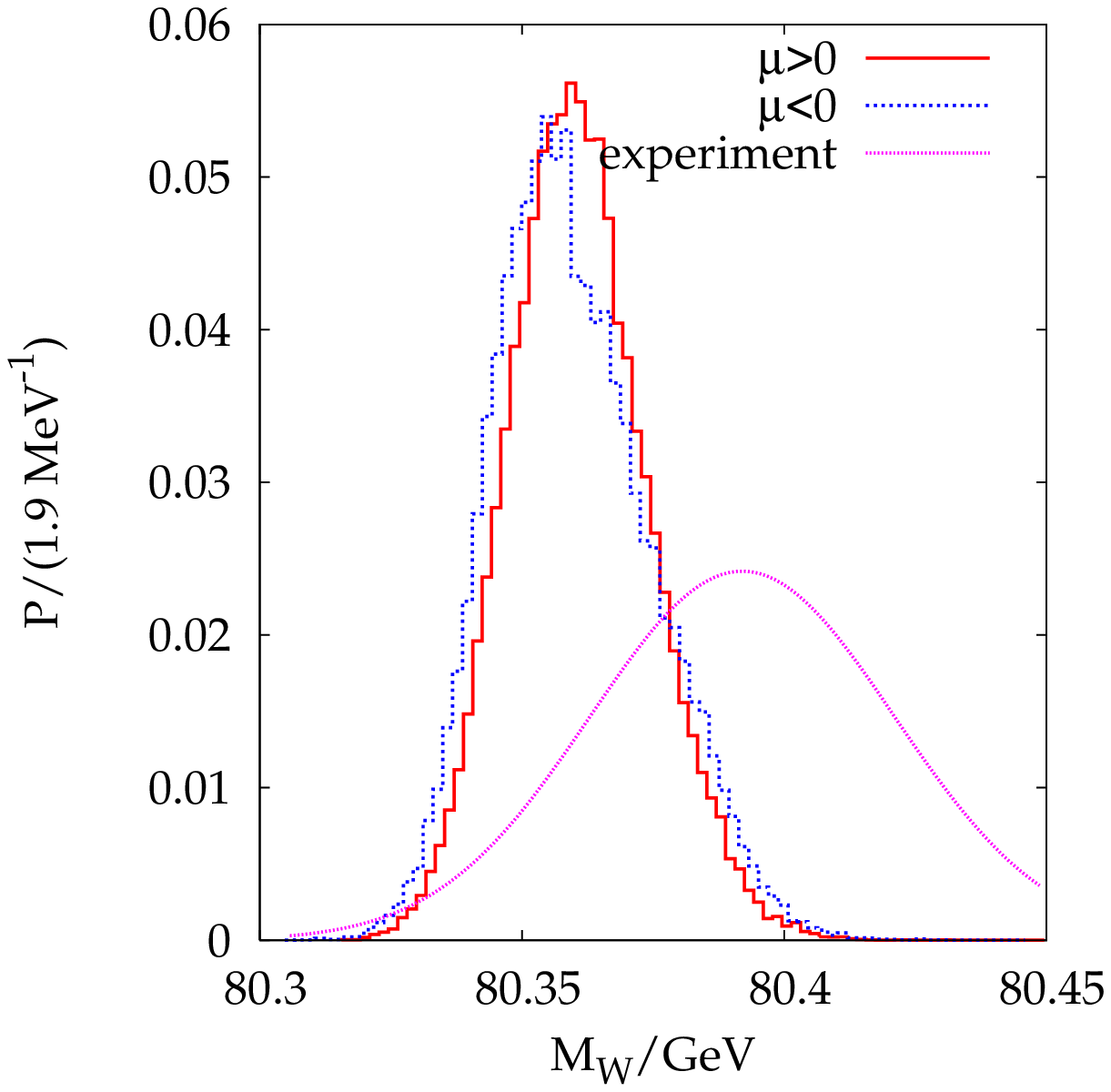}{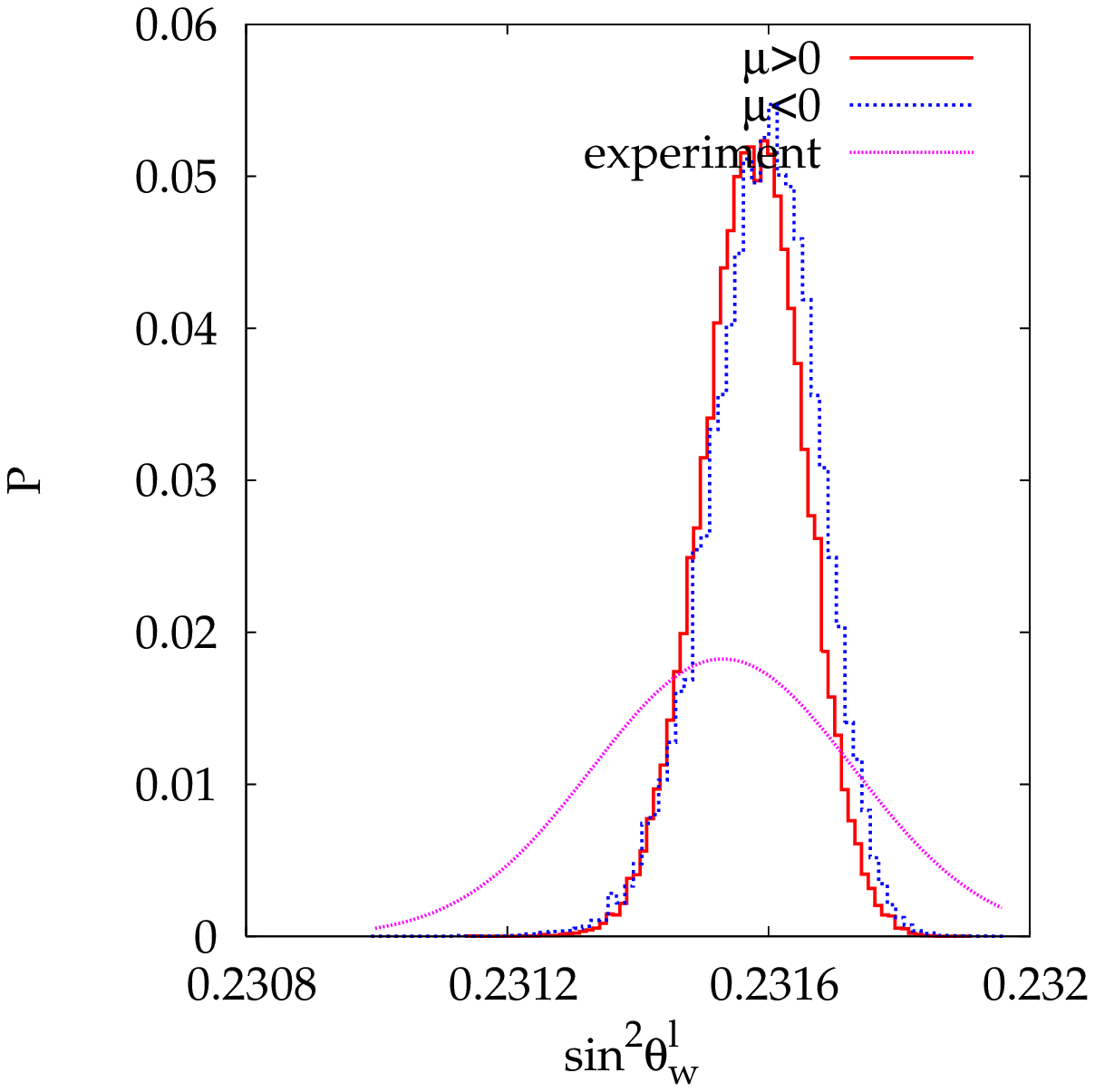}
\caption{Posterior probability distributions for weak observables in
  mSUGRA\@. \label{fig:weakobs}} } 
Figs.~\ref{fig:weakobs}(a) and~\ref{fig:weakobs}(b) show the probability
  distributions for $M_W$, 
  $\sin^2 \theta_w^l$ coming from the mSUGRA fits for both signs of
  $\mu$, although the sign does not make much difference.  We see that
  the $M_W$ prediction coming from the fits is skewed towards values
  lower than the central empirical value, corresponding to a
  preference for heavy SUSY from the rest of the
  fits. Fig.~\ref{fig:mw} confirms that heavier SUSY tends to have
  lighter values of $M_W$.  From Fig.~\ref{fig:st}, we see that heavy
  SUSY tends to be on the upper 1$\sigma$ empirical limit of $\sin^2
  \theta_w^l$.  Fig.~\ref{fig:weakobs}(b) does show evidence for this
  skew, which is rather mild.
\DOUBLEFIGURE{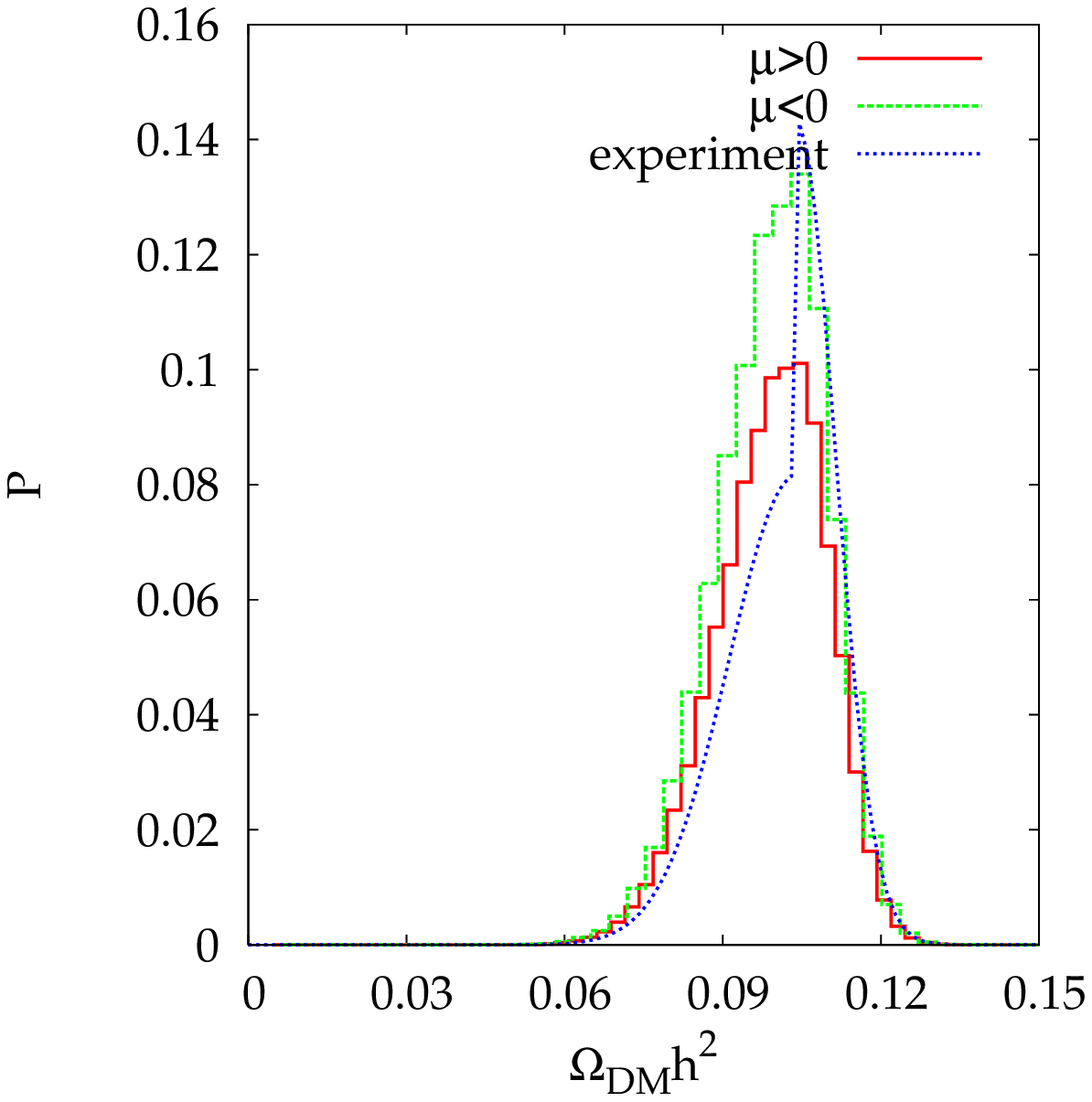,width=7cm}{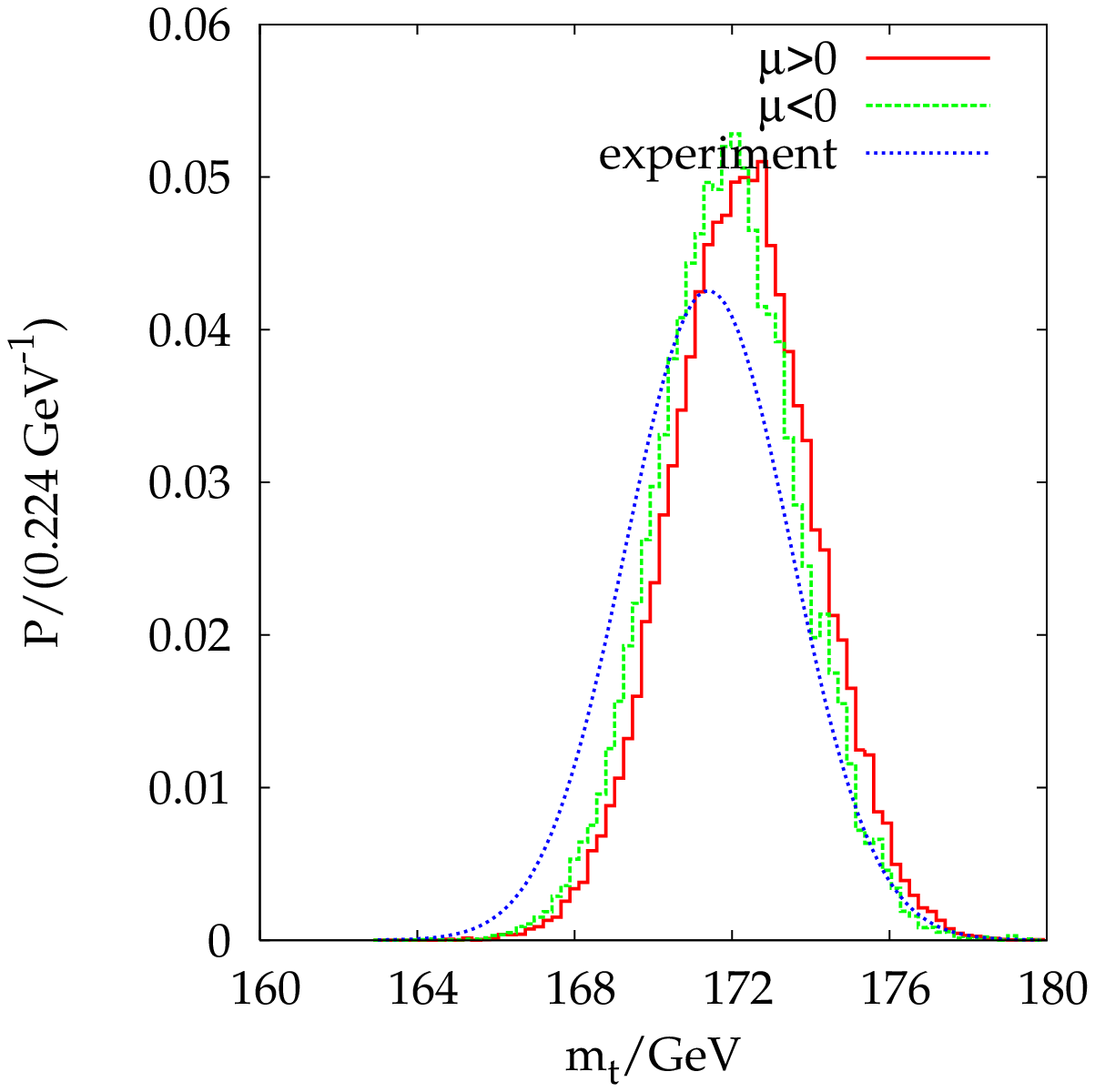,width=7cm}{Dark
  matter relic density probability distribution \label{fig:whsq}}{Top
  mass probability distribution
  \label{fig:mt}} 

\EPSFIGURE{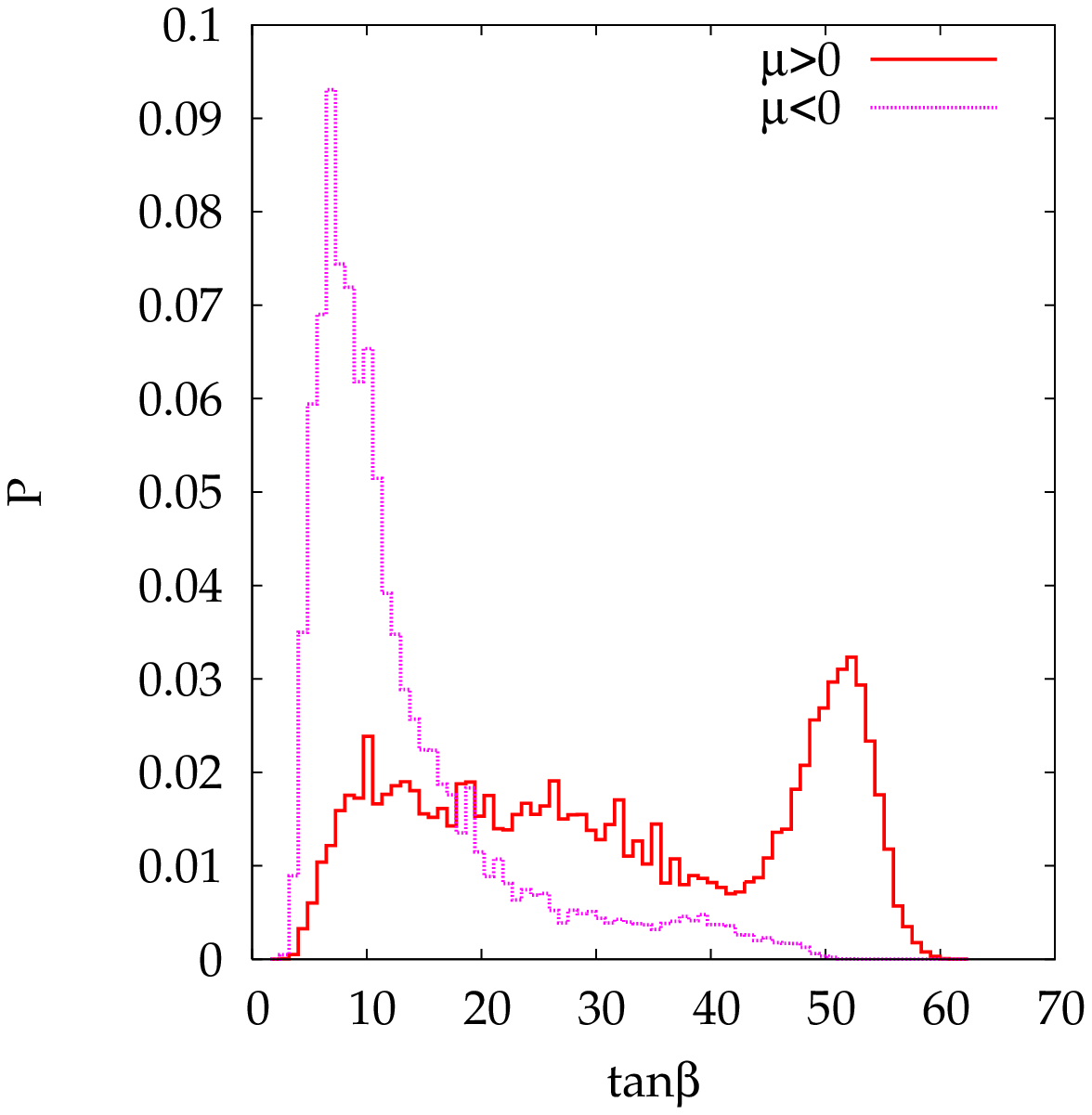,width=7cm}{Probability distribution for $\tan \beta$ in mSUGRA
\label{fig:tb}}
The strongest constraint in the fits comes from the dark matter relic
density.  In Fig.~\ref{fig:whsq}, we show the probability densities
resulting from the fits for the dark matter relic density. Each curve
is normalised slightly differently to allow better viewing of the
results. We see that both the $\mu>0$ and the $\mu<0$ $\Omega_{DM}
h^2$ distributions follow the empirical constraint closely, except for
a slight excess just lower than the central value.

Another important aspect influencing the fits is the integrated {\em
volume} \/of the probability density, which is automatically taken
into account in a Bayesian analysis, as was demonstrated and pointed out in
Ref.~\cite{deAustri:2006pe}.  The usual arguments based purely on
values of $\chi^2$ above the best-fit value are valid when the probability
density function is Gaussian in the interesting parameters
(which is certainly not the case here, as even a
cursory glance at Fig.~\ref{fig:m0m12} allows). Thus even if, say, the
stop co-annihilation region had a much lower $\chi^2$ than other
regions of the fits,\footnote{In reality, the stop co-annihilation
region fits the data rather marginally.} the fact that its volume in
8-dimensional input parameter space is much smaller than the other
regions will automatically disfavour it since its integrated
probability will be low.

The LEP2 higgs constraints shown in Fig.~\ref{fig:mhpen}, along with
the current empirical value of $m_t$ shown in Table~\ref{tab:inp}
favours rather heavy mSUGRA\@.  Indeed, Fig.~\ref{fig:mt} shows that
$m_t$ is skewed to somewhat higher masses than the empirical
constraint, illustrating this tension (again, we have altered the
normalisation of the curves slightly for clarity).  Inspection of the
$\alpha(M_Z)^{\overline{MS}}$, $\alpha_s(M_Z)^{\overline{MS}}$,
$m_b(m_b)$ inputs show that they follow their empirical probability
distributions very closely.  There is a large volume of parameter
space for the $A^0$ dark matter annihilation region at high $\tan
\beta>10$ particularly for $\mu>0$, as shown in Fig.~\ref{fig:tb}.
High values of $\tan \beta$ mean that light SUSY is disfavoured by
$BR[b \rightarrow s \gamma]$ and $BR[B_s \rightarrow \mu^+ \mu^-]$
since the data disfavours SUSY contributions~\cite{Ellis:2005sc},
which are approximately proportional to $\tan^2 \beta / M_{SUSY}^4$
and $\tan^6 \beta / M_{SUSY}^4$ respectively.  The distributions of
these two observables are shown in Fig.~\ref{fig:bs}.  
As can be seen
from the figure, the sign of the SUSY contribution to each observable
depends upon the sign of $\mu$.  The maxima in each figure correspond
to observables being close to their SM limit. 
$BR[b   \rightarrow s \gamma]$ prefers $\mu>0$ mildly, as $\mu<0$ tends to
predict a $BR[b   \rightarrow s \gamma]$ larger than the central empirical
value. 
\FIGURE{\twographs{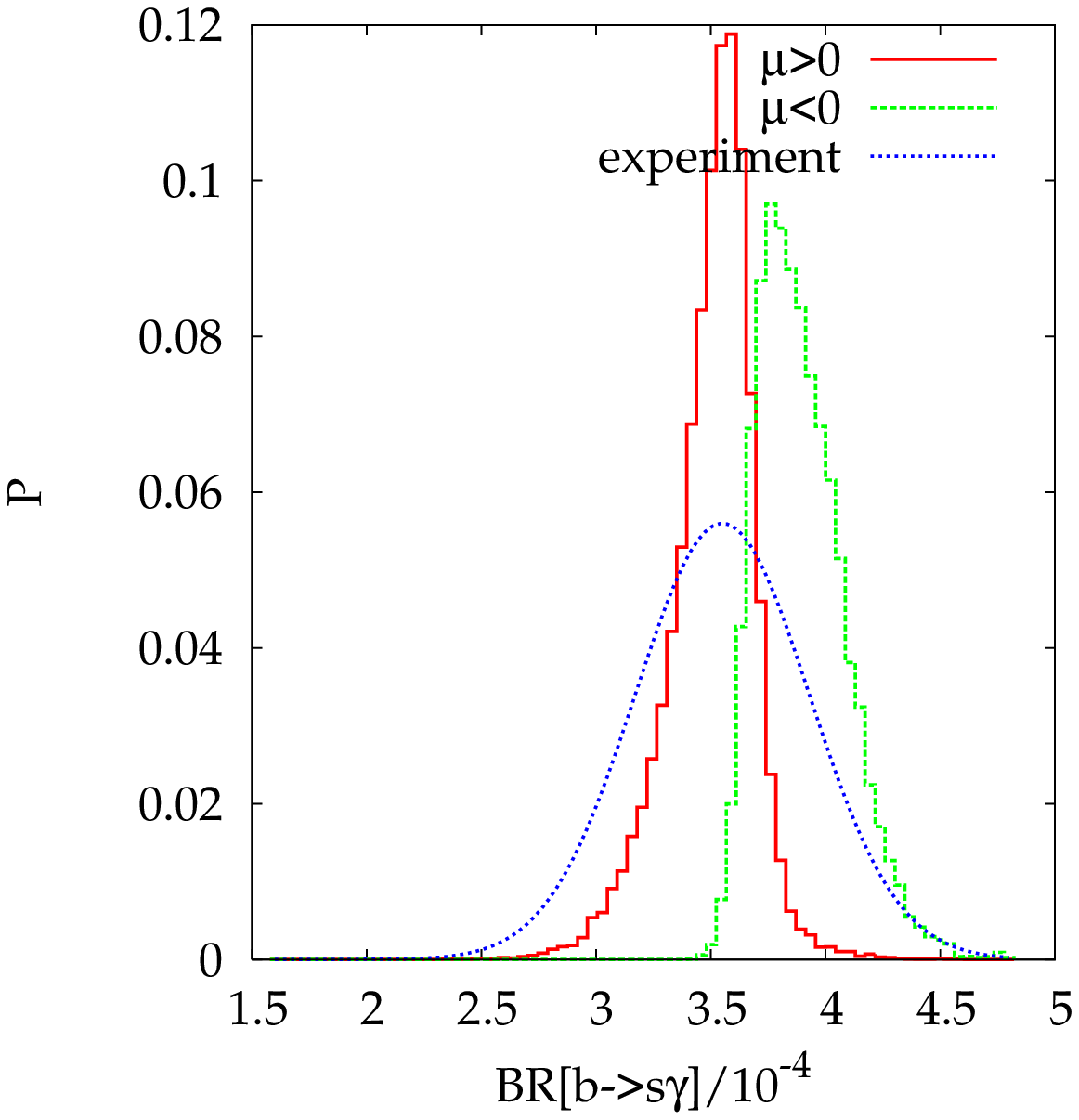}{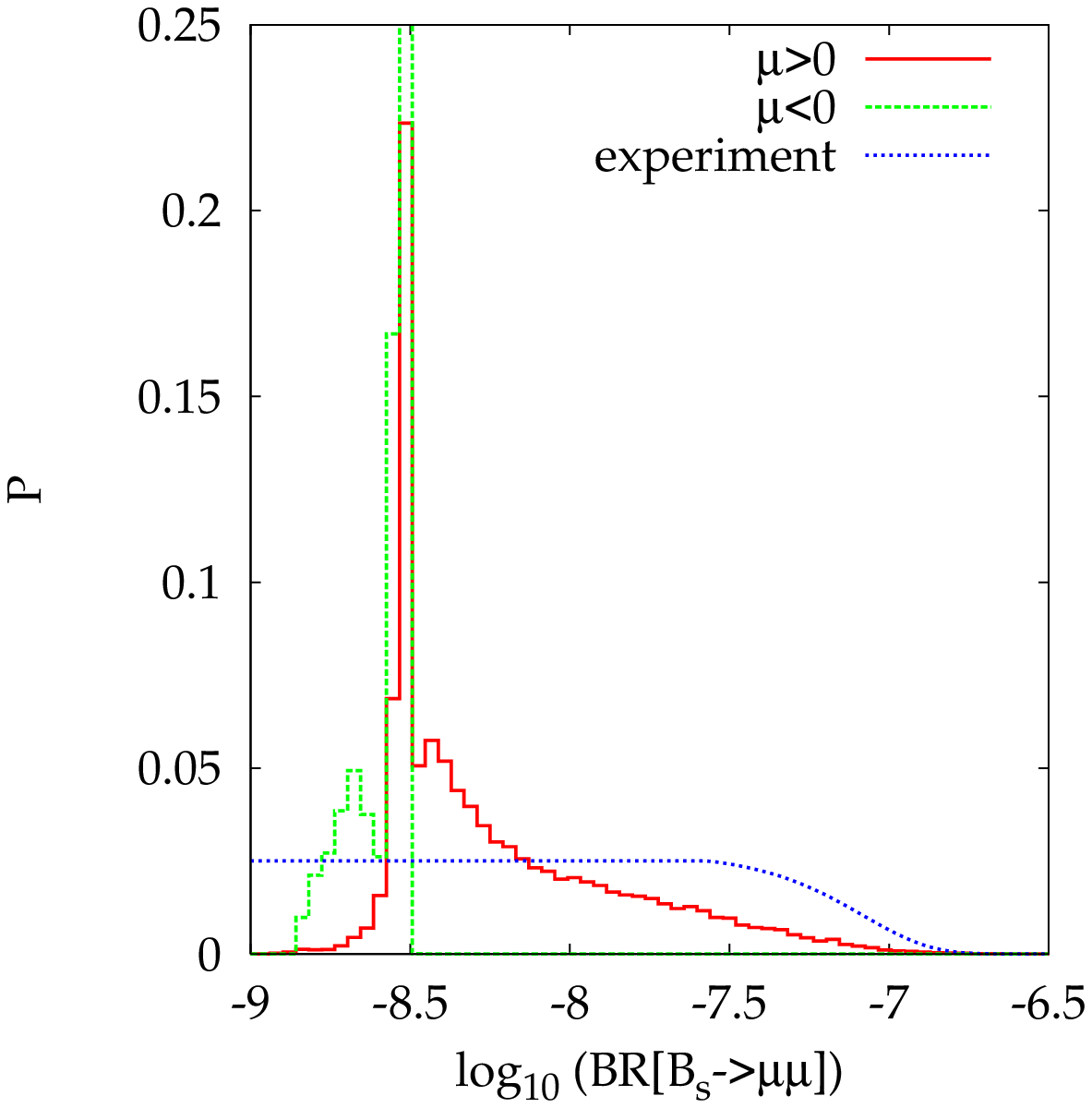}
\caption{Probability distributions of (a) $BR(b \rightarrow s
    \gamma)$ and (b)$BR(B_s \rightarrow \mu^+ \mu^-)$ in
    mSUGRA\@. \label{fig:bs}} 
}

\FIGURE{\twographs{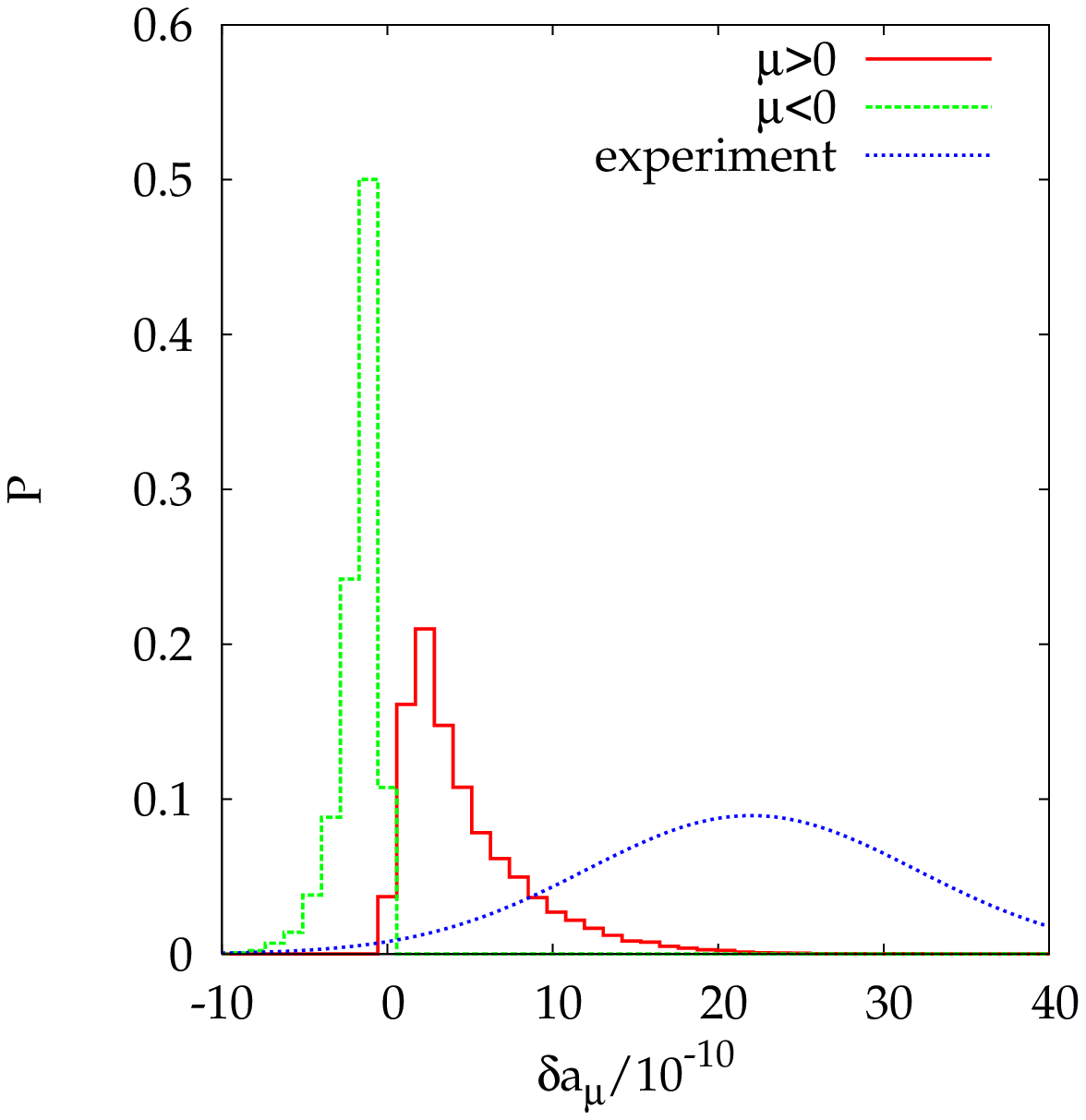}{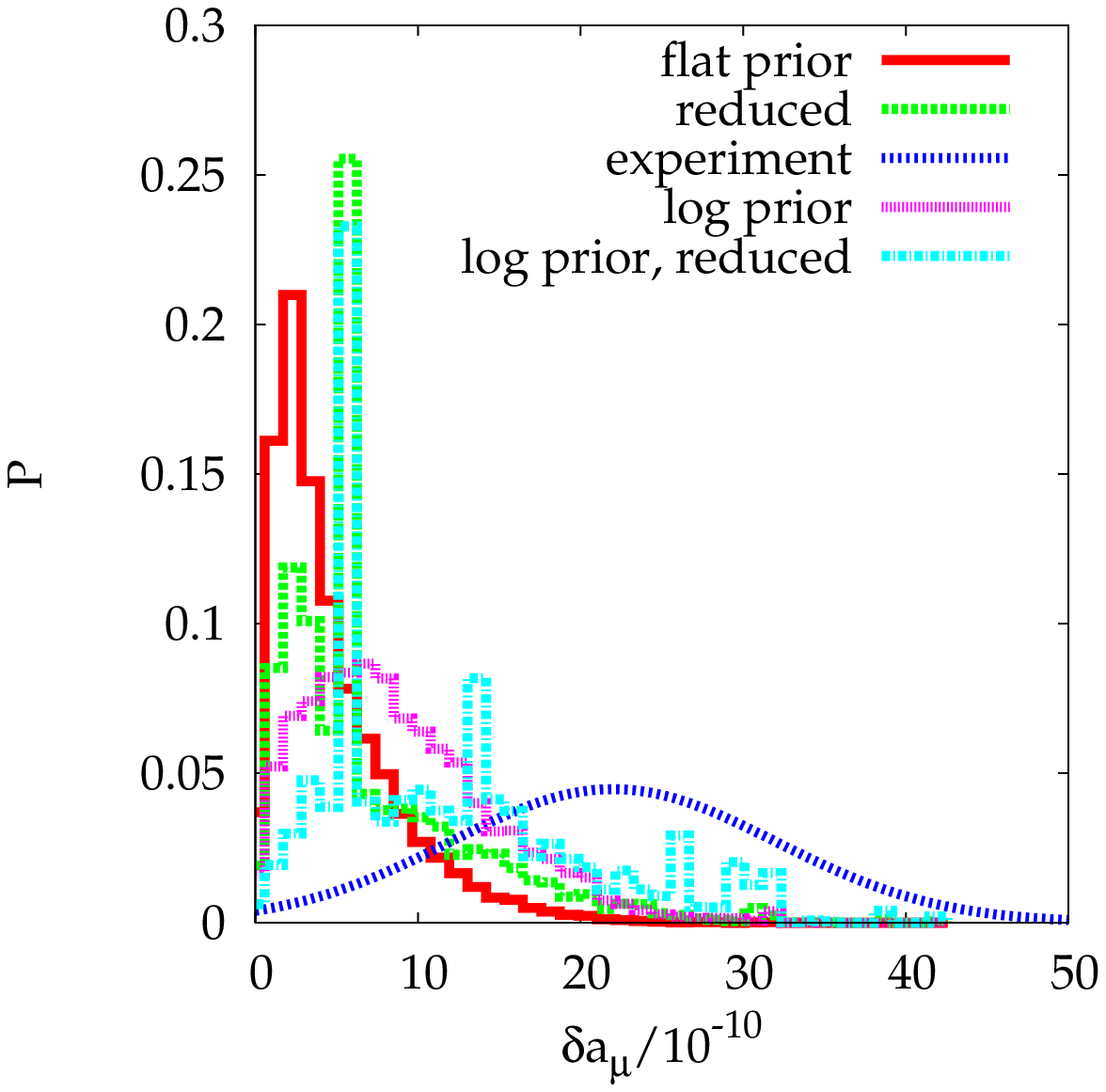}
\caption{Probability distributions for $\delta a_\mu$ in mSUGRA (a)
  for flat priors and both signs of $\mu$ and (b) for $\mu>0$ and
  various different priors. The key is explained in the
  text.\label{fig:gm2}} } $(g-2)_\mu$ is expected to be the observable
  that most strongly discriminates between the two signs of $\mu$. We
  plot its distribution in each case in Fig.~\ref{fig:gm2}(a). Since
  $\mu<0$ has the wrong sign of $\delta a_\mu$ compared to experiment,
  the probability density bunches around zero. Clearly heavy SUSY with
  a less negative SUSY contribution is favoured. However, we see that
  the $\mu>0$ distribution also prefers smaller SUSY contributions
  than the data in the global fits. This was initially unexpected, and
  will lead to $\mu<0$ mSUGRA being less ruled out.
In order to understand this behaviour better, we re-weight the $\mu>0$ sample
in various ways. $\delta a_\mu$ is approximately proportional to $\tan^2 \beta
  / M_{SUSY}^4$ and we need large $\tan \beta$ and rather light SUSY in order to
  get a sizable value in line with the central experimental value. As
  explained above, this is somewhat in conflict with LEP2 Higgs constraints
  and    $BR(b \rightarrow s \gamma)$. We re-weight the $\mu>0$ chains,
  dividing by a 
  number that removes the likelihood contribution from the LEP2 Higgs
  constraint and the  $BR(b \rightarrow s \gamma)$ measurement, i.e.\
  ${\mathcal L}_h {\mathcal L}_{BR(b \rightarrow s \gamma)}$. The probability
  distribution of $\delta a_\mu$ resulting from this procedure  is marked 
  as the ``reduced'' curve in Fig.~\ref{fig:gm2}(b). It extends to somewhat
  higher and more central values of $\delta a_\mu$, showing that some of the
  skew in the $\delta a_\mu$ distribution does come from the LEP2 Higgs and
  $BR(b \rightarrow s \gamma)$ measurements. However, the effect is rather
  small and the resulting distribution is rather far from the experimental
  distribution, indicating a further effect. There could be a significant
  volume effect if regions of parameter space that have central values
  of   $\delta a_\mu$ have a small volume. This would render our results
  sensitive to the prior, since by changing the measure of the input
  parameters, we can change the volume measure~\cite{Allanach:2006jc}.
In order to investigate the effects of this, we re-weight the $\mu>0$ chains
by a factor $1/(m_0 M_{1/2})$. Such a re-weighting mimics the effect of using
  a logarithmic prior on $m_0$ and $M_{1/2}$ since
\begin{eqnarray}
\int P(m_0, M_{1/2})\  d \ln m_0\ d \ln M_{1/2} &=& \int P(m_0, M_{1/2}) \frac{d \ln m_0}{d m_0}
\frac{d \ln M_{1/2}}{d M_{1/2}} \ dm_0 \ dM_{1/2} \nonumber \\
& = & \int \left(\frac{P(m_0,
  M_{1/2})}{m_0 M_{1/2}} \right) \ dm_0 \ dM_{1/2}. \label{rewt}
\end{eqnarray}
One might consider such a prior on technical naturalness grounds. 
The ``log prior'' results are displayed in Fig.~\ref{fig:gm2}(b). They have a
fatter tail out to higher values of $\delta a_\mu$ than the flat prior sample
or the ``reduced'' sample. When we perform the re-weighting in Eq.~\ref{rewt}
as well as the one to remove $BR(b \rightarrow s \gamma)$ and LEP2 Higgs
constraints, we obtain the ``log prior, reduced'' curve in the figure.
This shows a yet fatter tail and is starting to approach the empirical
probability distribution imposed on the fits. 
Our results are obviously somewhat dependent upon the prior. This is
essentially because there is not enough precise data yet to constrain 
mSUGRA very strongly. We must bear in mind this dependence upon the prior, and 
investigate different priors when we estimate $P_-/P_+$ below. 

\TABULAR{|c|ccc|}{\hline
prior &  flat & small & log \\ \hline
$P_-/P_+$ & 0.16 & 0.12 & 0.07\\
\hline
}{Ratios of integrated probability for different signs of $\mu$.
  \label{tab:ratios}}
The total normalisation of the $\mu>0$, $\mu<0$ samples is 
shown in
Table~\ref{tab:ratios} for various different priors. 
The ``small'' prior is a flat prior with a reduced range compared to the one
displayed in Table~\ref{tab:inp}. We filter the chains to only include points
with $m_0<2$ TeV and $|A_0|<2$ TeV. The table shows that
$\mu<0$ is somewhat disfavoured for the smaller ranges and more disfavoured
still for the log prior. The result is somewhat sensitive to the prior,
indicating the need for more data. We therefore prefer to quote a range 
for $P_-/P_+=0.07-0.16$ depending upon the prior. This range is a focal result
of the present paper.

\section{Fits Without WMAP3 \label{sec:noDM}}

\FIGURE{\twographst{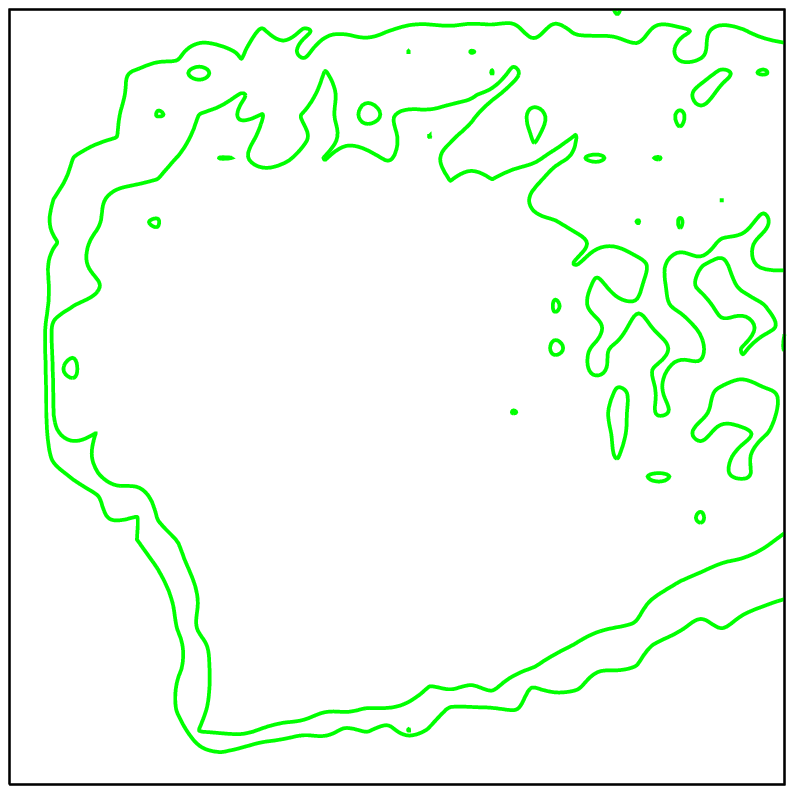}{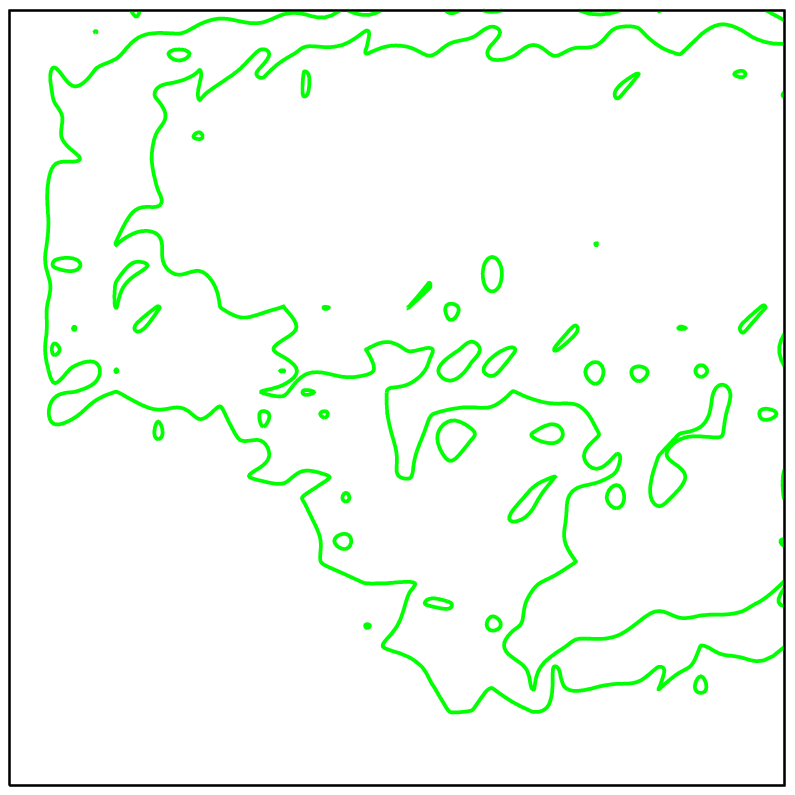}
\caption{Constraints from global fits {\em without the dark matter constraint}
  \/with (a) $\mu>0$ and (b) $\mu<0$ mSUGRA 
  marginalised to the $m_0$-$M_{1/2}$ plane.
We have assumed a flat prior. The posterior probability is indicated by the bar
  on the right hand side. The inner (outer)  contours show the
  boundary of a $68\%$
  $(95\%)$ confidence region. \label{fig:noDM}
}
}
We now briefly examine the effect of removing the dark matter $\chi^2$ penalty
from the fits. We could in principle re-weight the chains in order to do this,
but we find that that leads to statistical fluctuations in the results that
are too large. Initial investigations revealed that the efficiency becomes
much higher when we remove the dark matter relic density contribution to the
total $\chi^2$. We are able to increase the widths of the proposal
distribution to 197 GeV for $m_0$, 100 GeV for $M_{1/2}$, 400 GeV for $A_0$,
8.5 for $\tan 
\beta$ and 1$\sigma$ for the Standard Model inputs and {\em still} \/achieve an
efficiency of around 35$\%$. This has the consequence that the chains explore
the parameter space much quicker than in the previous section, and so less
MCMC steps are needed.
We run a further 9$\times$200~000 MCMC steps for each sign of $\mu$. This
time, both signs of $\mu$ have excellent convergence statistics, $\hat R$
being different to 1 at the per-mille level only in each case.
\FIGURE{\fourgraphs{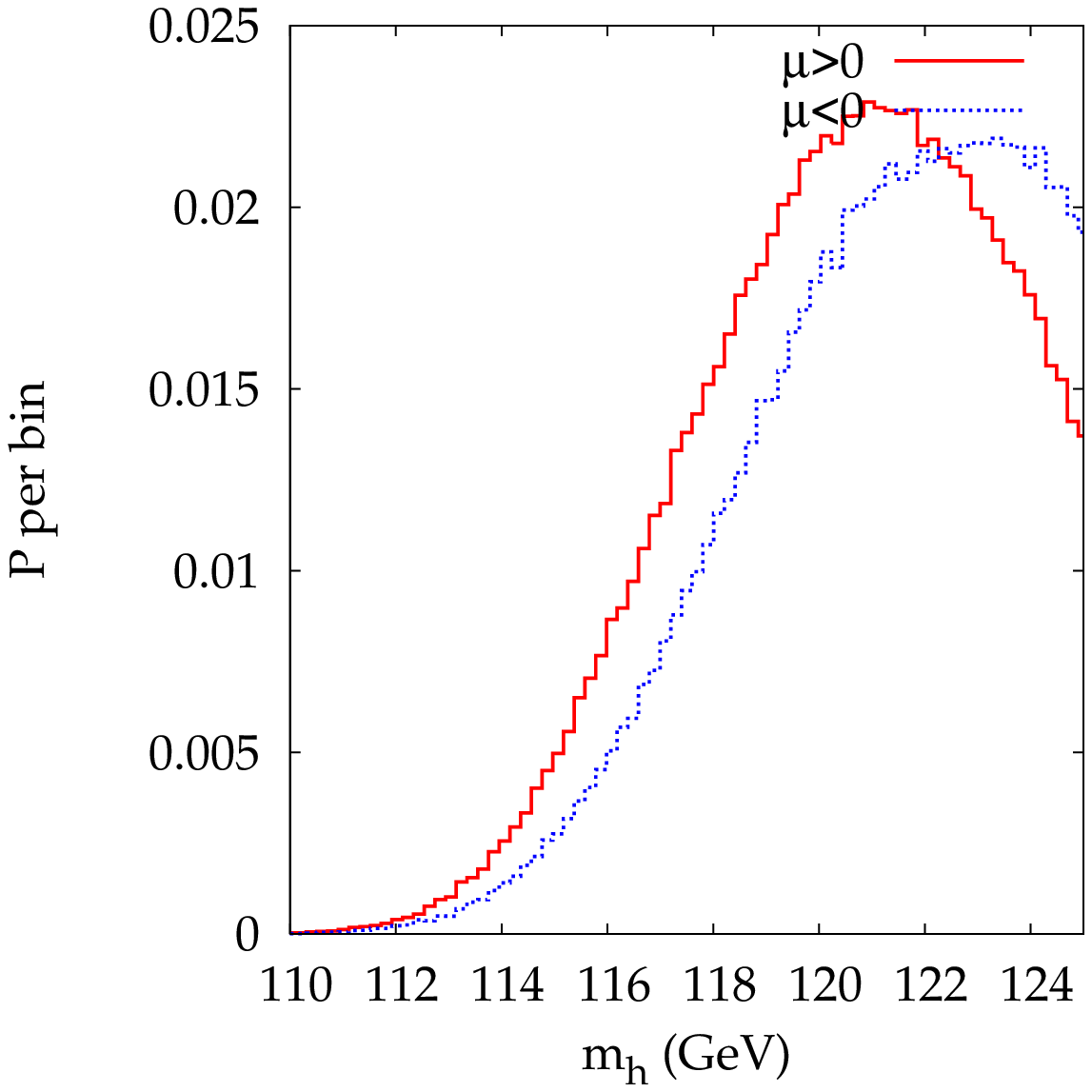}{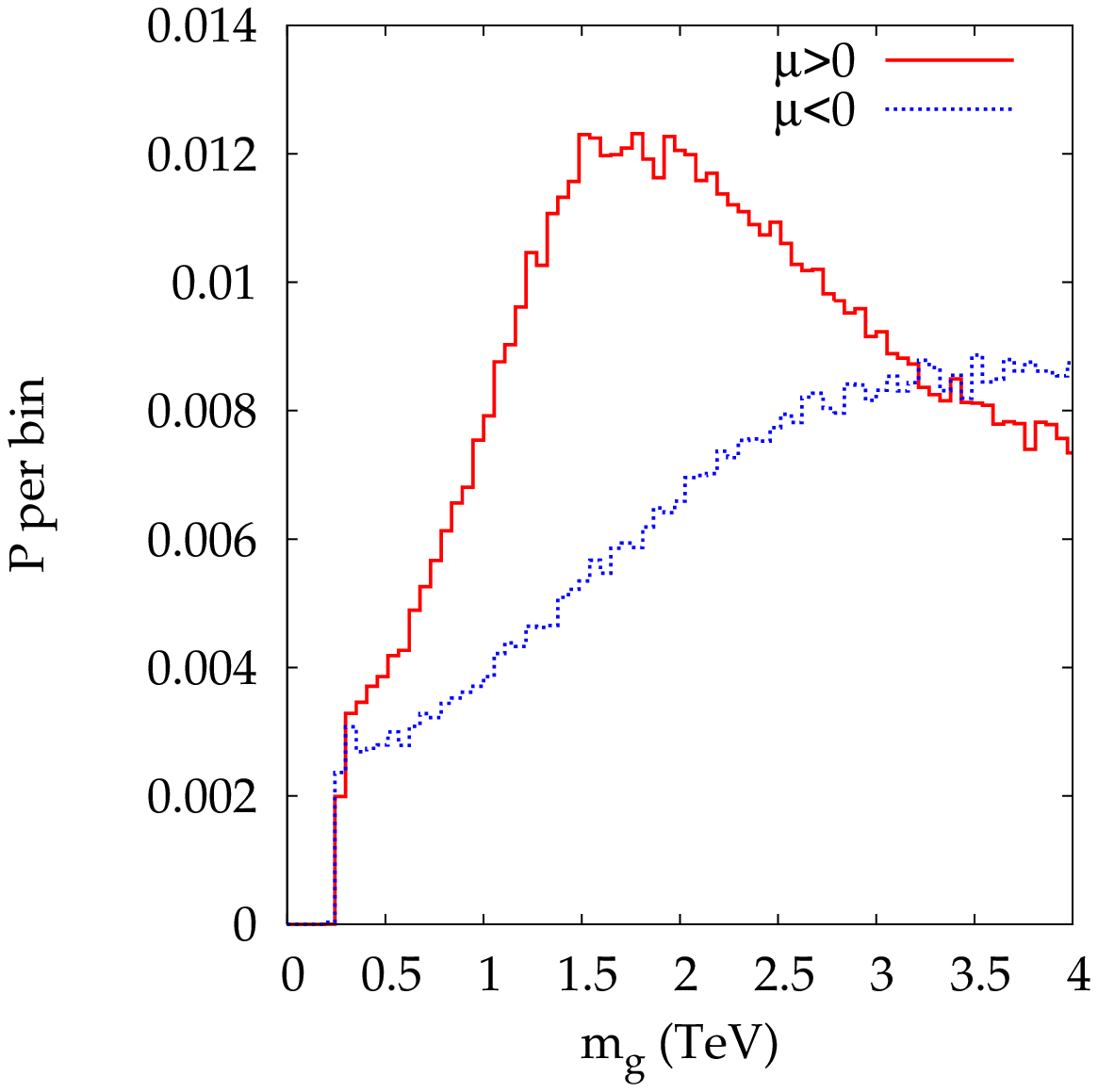}{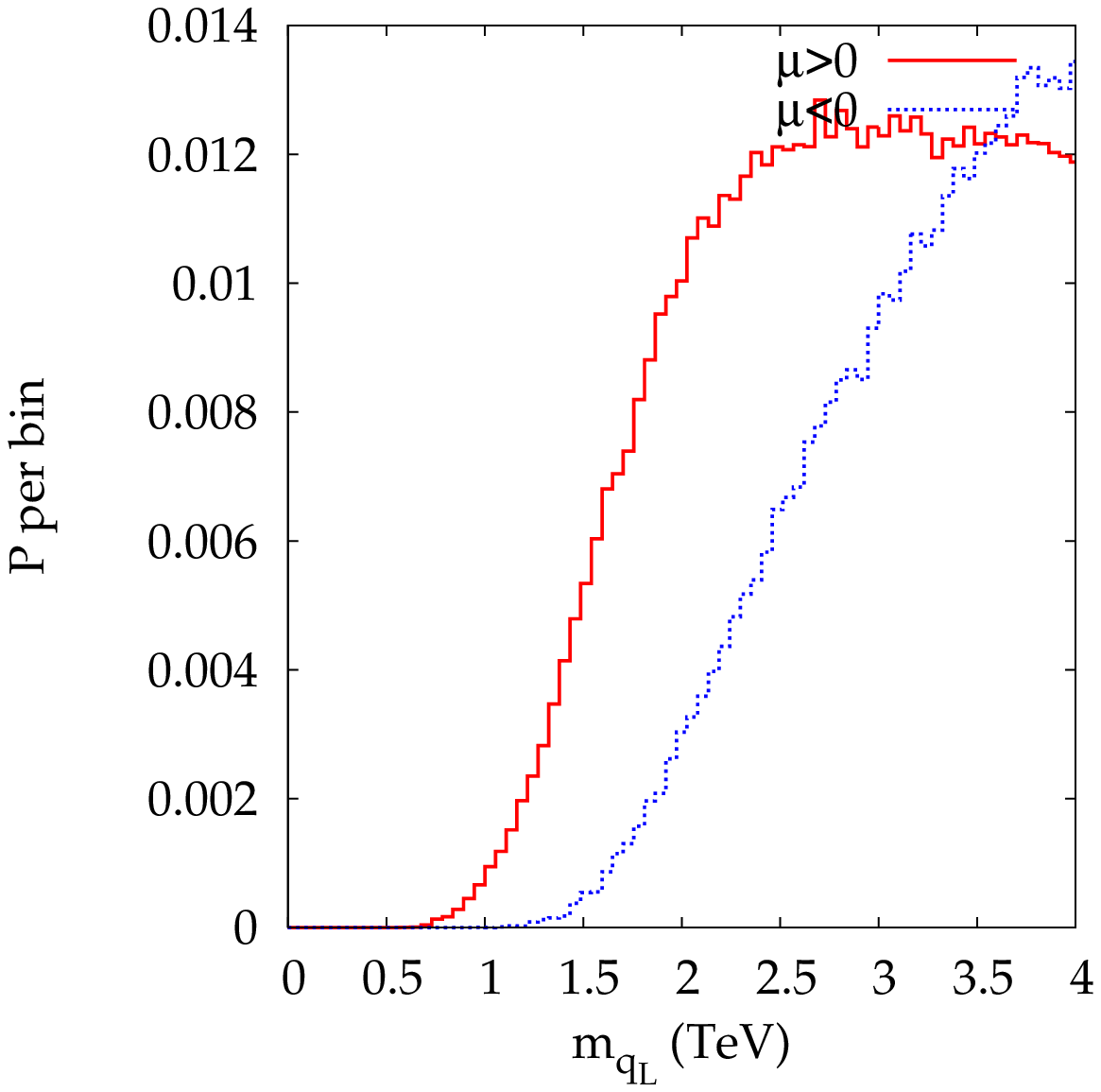}{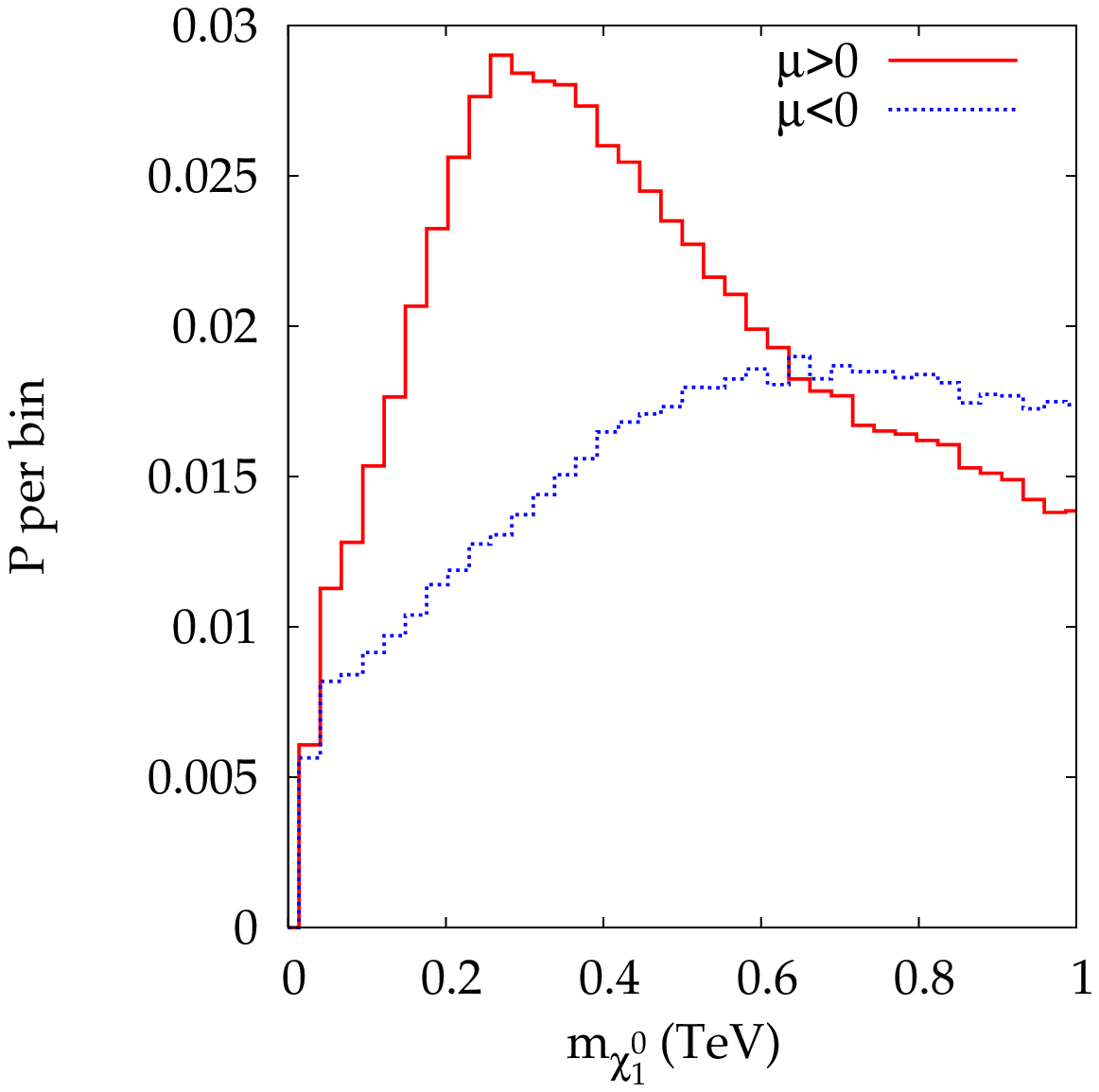}
\caption{Probability distributions in mass of (a) the lightest CP even higgs,
 (b) gluino, (c) the left-handed squark and (d) the neutralino. Flat priors
 have been assumed.
\label{fig:massNDM}}
}

\EPSFIGURE{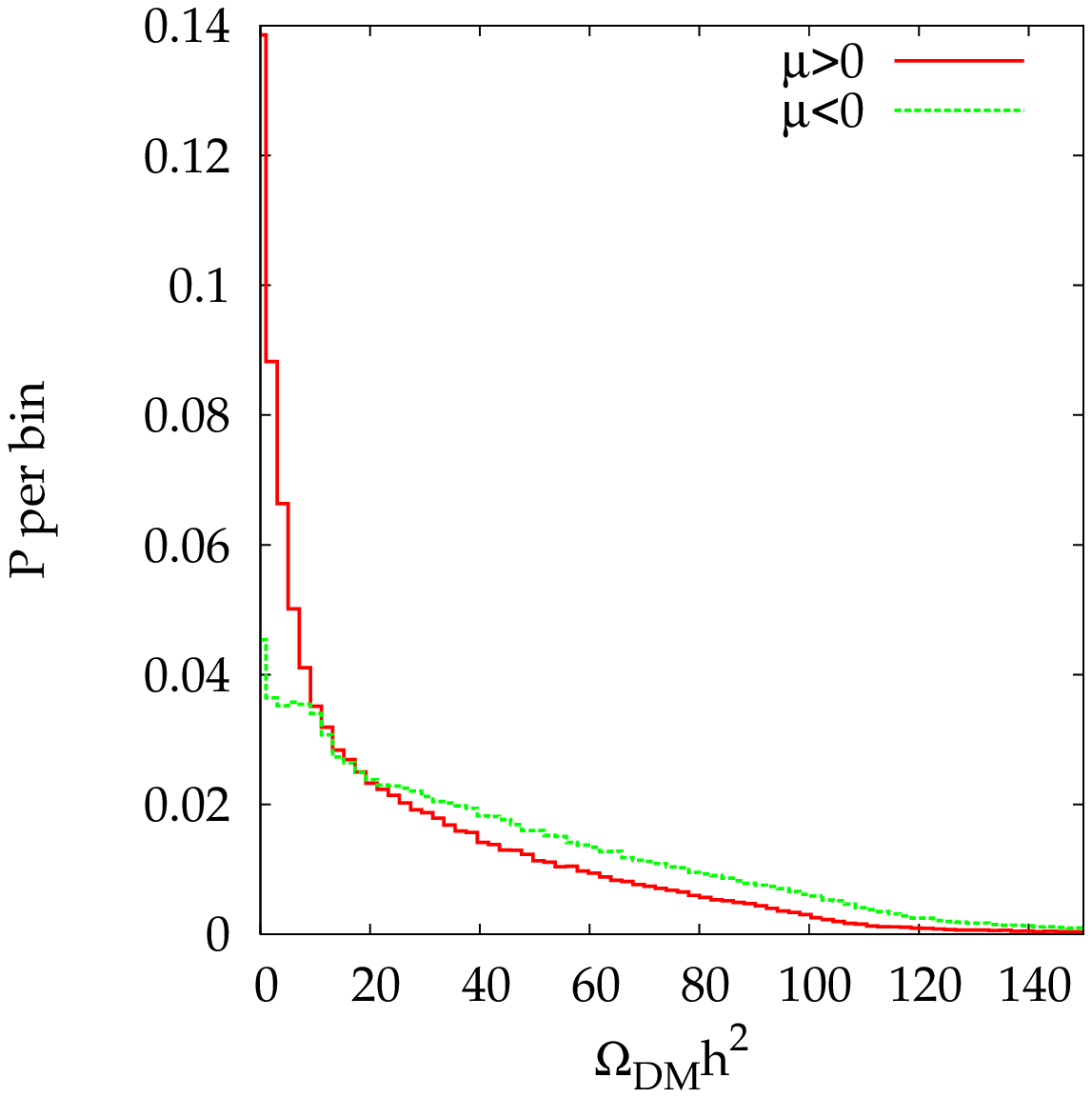,width=6.4cm}{$\Omega h^2$ distributions without imposing the
dark matter constraint. \label{fig:om}}
We see from Fig.~\ref{fig:noDM} that removing the dark matter constraint
  yields a very different picture for the mSUGRA probability distributions in
  the $m_0$-$M_{1/2}$ plane. This confirms our statement that many of the
  features seen in the previous section 
  are due precisely to that constraint. The probability distribution is now
  much flatter in the input parameter space.  
The disallowed region at low $m_0$ and high $M_{1/2}$ is due to the no-charged
  LSP constraint. The disallowed region at low $m_0$ and low $M_{1/2}$ is due
  mostly to the combined effect of the $(g-2)_\mu$, $BR(b\rightarrow s \gamma)$
  and $m_h$ constraints. The disallowed region is significantly larger for
  $\mu<0$ than for $\mu>0$, due to those three observables. Marginalisations
  in other mSUGRA input parameter planes tell a similar
  story, more featureless than the fits including the dark matter constraint. 
As mentioned in the introduction, Fig.~\ref{fig:noDM} covers the case of
  R-parity violating mSUGRA when the R-parity violating couplings are smaller
  than about
  0.1. For larger R-parity violating couplings, one would have to include
  them in the renormalisation group equations to obtain accurate results.

The more featureless fits have predictable effects on the mass spectrum of the
MSSM, as shown in Fig.~\ref{fig:massNDM}. We have fixed the ranges of the
abscissas to be identical to those in Fig.~\ref{fig:mass} in order to
facilitate comparison. The lightest CP-even Higgs probability distribution is
broader and shifted to heavier values, due to the bigger volume of parameter
space allowed at higher $m_0$ and $M_{1/2}$ values. Gluino mass distributions
no longer tail off at higher masses: upper bounds would be mostly determined
by the cut-off placed $m_0$ and $M_{1/2}$. The gluino mass and neutralino
distributions are 
also much less peaked than the ones in Figs.~\ref{fig:mass}b,d particularly for
$\mu<0$. The flatter distributions are, of course, an indication that the data
aren't strongly constraining. Large volumes at large $m_0,M_{1/2}$ effectively
move up the squark masses, as Fig.~\ref{fig:massNDM}c illustrates.

The probability distribution of $\Omega_{DM} h^2$ is shown in
Fig.~\ref{fig:om} for each sign of $\mu$, when we drop the relic dark matter
$\chi^2$ contribution. In the figure, we see that huge
values can result: in fact the mean value for the $\mu>0$ sample
tails off at $\Omega_{DM} h^2 \sim 128$, much
larger than the WMAP3 value of 0.1. 
\TABULAR{|c|ccc|}{\hline
prior& flat & small  & log  \\ \hline
 ${P_-/P_+}$ &
  {0.45} &
 {0.43} &
 {0.19} \\
\hline
}{Ratios of integrated probability for different signs of $\mu$ with no dark
 matter constraint.
  \label{tab:ratios2}}
This illustrates the fact that regions of parameter space which fit the dark
matter data in mSUGRA need some special annihilation mechanism that is not
typical of the whole space. This and similar arguments have led several
authors to consider non-universal models~\cite{gian,sfk}, where the relic
density might be less fine-tuned. The distributions are highly skewed, having
tiny tails up to $\Omega h^2 \sim 1000$. 
The relative normalisation of $\mu>0$ to $\mu<0$ is calculated in the same
manner as in the previous section for different priors, and displayed in
Table~\ref{tab:ratios2}. 
$\mu<0$ is hardly disfavoured in R-parity violating mSUGRA (where we can
neglect the dark matter constraint), where
$P_-/P_+=0.19-0.45$.

\section{Conclusions \label{sec:conc}}

We have performed global fits to mSUGRA using indirect data and
state-of-the-art predictions of the observables. The MCMC technique was
successfully employed despite initial non-convergence of the $\mu<0$ chains.
Bridge sampling was used to normalise two isolated maxima that had not been
traversed by any chain for $\mu<0$. 
We found that $\mu<0$ is somewhat disfavoured in comparison to $\mu>0$ but not
by huge margins. The rest of the fit prefers rather heavy SUSY and so the SUSY
contribution to $(g-2)_\mu$ is small whichever the sign of $\mu$. 
$\mu<0$ is only disfavoured marginally, 
the ratio of integrated probability densities being
$P_-/P_+=0.07-0.16$ depending upon the prior.
We see from Fig.~\ref{fig:om} that without
the dark matter constraint, $\Omega_{DM} h^2$ is predicted up to values of
around 128. This corresponds to a $\chi^2_{\Omega}$ of around 3$\times 10^8$,
much larger than is likely from the other observables such as $M_W$ or 
$\sin^2 \theta_w^l(\mbox{eff})$.  
The fits are therefore completely dominated by the
dark matter relic density constraint and volume effects. 
Expectations that the SUSY scale will be light because of a preference from
weak observables turn out not to be true in the global fits. 
If the dark matter constraint is dropped, as would be the
case for R-parity violation, $\mu<0$ is hardly disfavoured at all, 
$P_-/P_+=0.19-0.45$. $\mu<0$ is much less disfavoured than many seem to
assume.  
Many recent analyses only consider $\mu>0$ on the grounds that $\mu<0$ is
strongly 
disfavoured by $(g-2)_\mu$. We have therefore demonstrated that this is {\em
  not} \/the case when one considers the entirety of the data and that $\mu<0$ 
should still  be considered in mSUGRA analyses.

It could be argued that the flat measure used here in $m_0$, $M_{1/2}$, $A_0$
and $\tan \beta$ can be improved upon. For instance, $\tan \beta$ is really
a derived quantity and is related to more fundamental Higgs potential
parameters, which could be considered more natural to have a flat measure
upon. There is also the issue of fine-tuning, recently illuminated in
Ref.~\cite{Giudice:2006sn}: we could disfavour regions of parameter space that
are highly fine-tuned, for instance~\cite{Allanach:2006jc}. Changes such as
these in the prior could potentially change the results of the fits and we
intend to investigate them in a future publication. 

Clearly, more data is required to 
decrease the dependence of results on the prior.
The most helpful data is likely to be that from colliders. 
The MCMC fitting technique has been used in an ATLAS study examining how
cross-section and kinematic endpoint information constrains
mSUGRA and non-universal models~\cite{Lester:2005je}.
At this moment, without data from colliders, we are forced to use indirect
constraints for the observables. However, in the future it will be desirable
to predict $\Omega_{DM} h^2$ given some SUSY collider
observables~\cite{Allanach:2004xn,Baltz:2006fm}. If this is in contradiction
with the observed value from cosmology, it will point to a wrong cosmological
assumption, which could then be changed. In order to really confirm that 
dark matter particles have been produced at colliders, one requires
compatibility with direct dark matter detection data. 
Of course one would like to drop the mSUGRA assumption and perform a general
SUSY analysis, but for this it is likely that additional data from a future
international linear collider would be
required~\cite{Allanach:2004xn,Baltz:2006fm}.  
In any case, the techniques investigated in this paper should prove useful for
the fits.

\acknowledgments
This work has been partially supported by PPARC\@. This paper was produced
using the University of Cambridge EScience CAMGRID computing facility.
We would like to thank W Hollik for help with electroweak observables, 
B Heinemann and C S Lin for the likelihood density penalty of $B_s \rightarrow
\mu^+ \mu^-$, R Rattazzi for a discussion about priors,
D St\"ockinger for the two-loop contributions to the anomalous magnetic
moment of the muon, A Pukhov for help with {\tt micrOMEGAs},
J Ellis, K Olive, T Plehn, L Roszkowski, J Smillie, G Weiglein and the
Cambridge SUSY 
working group for helpful comments and M Calleja for invaluable help with using
CAMGRID. 

\appendix

\section{Markov Chain Monte Carlos\label{sec:bridge}}

Our Markov chain consists of a list of parameter points (${\mathbf
x}^{(t)}$) and associated likelihood densities (${\mathcal L}({\mathbf x}^{(t)})$). Here, $t$ labels the link
number in the chain.  Given some point at the end of the Markov chain
(${\mathbf x}^{(t)}$), the Metropolis-Hastings
algorithm~\cite{Metropolis,Hastings,MacKay} requires one to randomly
pick another potential point (${\mathbf x}$) (typically in the
vicinity of ${\mathbf x}^{(t)}$) using a proposal distribution
$Q({\mathbf x};{\mathbf x}^{(t)})$. There is a large amount of freedom
in the choice of the proposal function $Q$, and this freedom is
usually exploited to improve the efficiency of the sampling process.
In order to ensure that the choice of $Q$ does not bias the final set
of samples in some way, the form of $Q$ is taken into account when
deciding whether to accept or reject the new point.  
If the ratio $\rho$
defined by
\begin{equation}
\rho = 
\frac{{\mathcal L}({\bf x})}{{\mathcal L}{({\bf x}^{(t)} )}}
\frac{Q({\mathbf x}^{(t)};{\mathbf x})}{Q({\mathbf x};{\mathbf x}^{(t)})}
\end{equation} is greater
than one, the new point ${\bf x}$ is appended to the chain.  If $\rho$
is instead less than one, a decision must be made to determine whether
to accept or reject the proposed point ${\bf x}$.  The rule is that
acceptance of ${\bf x}$ must occur with probability $\rho$.  If
accepted, ${\bf x}$ is added to the end of the chain. If not accepted,
the point ${\mathbf x}^{(t)}$ is copied once more onto the end of the
chain.  Whichever point makes it on to the end of the chain is
thereafter known as ${\mathbf x}^{(t+1)}$.

As a result of following the above steps, the sampling density of
points in the chain becomes proportional to the density of the target
distribution (such as the posterior probability density, or the
likelihood when the prior is uniform) as the number of links goes to
infinity, under the circumstances described in Ref.~\cite{MacKay}.  The
Metropolis-Hastings MCMC algorithm is typically much more efficient
than a straightforward scan for the dimensionality of input parameter
space $D>3$; the number of required steps scales roughly linearly with
$D$ rather than as a power law.  We take the proposal function $Q$ to
be a product of Gaussian distributions along each dimension
$k=1,2,\ldots,D$ centred on the location of the current point along
that dimension, i.e.\ $x^{(t)}_k$:
\begin{equation}
Q({\mathbf x}; {\mathbf x}^{(t)}) = 
  \prod_{k=1}^D \frac{1}{\sqrt{2 \pi} l_k} e^{-(x_k -x_k^{(t)} )^2
  / 2 l_k^2},
\end{equation}
where $l_k$ denotes the width of the distribution along direction $k$. 
For the case where we include the dark matter relic density in the
calculation, 
we choose $l_{m_0}=100$ GeV, $l_{M_{1/2}}=50$ GeV, $l_{A_0}=400$ GeV and
  $l_{\tan \beta}=3$. 
For the Standard Model inputs, we choose $l_k = 8 \sigma_k / 20$.  We
discuss why these particular values were chosen in the next section.

In order to start the chain we follow the following procedure, which finds a
point at random in parameter space that is not a terrible fit to the data.
We pick some ${\mathbf y}^{(0)}$ at random in the mSUGRA parameter
space using a flat distribution for its probability density function (pdf). 
The Markov chain for ${\mathbf y}$ is evolved through 2000 steps.
We then set ${\mathbf x}^{(0)}={\mathbf y}^{(2000)}$,
continuing the Markov chain in ${\mathbf x}$ and discarding the ``burn-in''
chain ${\mathbf y}$. A reasonable-fit point is typically found long before
2000 iterations of the Markov chain.
We must make sure that we perform enough iterations after this point 
that the chain traverses the remaining viable parameter space. 
We will provide a convergence test to this effect.

\subsection{Efficiency}

The efficiency of a chain can be defined as ``the
number of links whose coordinates differ from those of their
predecessor in the chain'' divided by ``the total number of points in
the chain''.  There will always be a tension between efficiency and
convergence\footnote{For a discussion of convergence, see
appendix~\ref{sec:conv}.} in chains:
if the $l_k$ are set to be too
small, efficiency will increase but the chain will take too long to
achieve convergence whereas if they are too large, the efficiency will
be so small that the sampling will contain large statistical
fluctuations. 
In practice, the bulk of our posterior
probability density is contained in a very thin hyper-surface in the
8-dimensional input parameter space~\cite{Allanach:2005kz}. It is thin
because $\Omega_{DM} h^2$ varies very rapidly over mSUGRA space
compared to the high accuracy of the empirical constraint. With the
$l_k$ listed above, we found that efficiencies were at the per-mille
level, too small to achieve a statistically stable result in a
reasonable amount of CPU time.  To achieve a significantly larger
efficiency, we had to reduce $l_k$ to such a level that we lost
convergence because the chains had not traversed the viable parameter
space.  In order to counter this, we expanded the errors on
$\Omega_{DM} h^2$ to $\pm 0.02$ while calculating the likelihood
density in the chain. This artificially thickens the ``surface''
containing the bulk of the posterior probability density, increasing
the efficiency to much more reasonable values of around 5--7$\%$.  In
order to correct for this artificial thickening and re-impose the
required constraint of Eq.~\ref{omega}, it was therefore necessary to
re-weight each link of each chain at the end of the sampling.  Each
link is re-weighted by the ratio of the proper likelihood density
${\mathcal L}_P$ to the likelihood density with inflated errors
${\mathcal L}_I$:
\begin{equation}
\frac{{\mathcal L}_P}{{\mathcal L}_I} =
        \exp \left(
	  {-\frac{(c_{\Omega}-p_{\Omega})^2}{2\sigma^2_{\Omega
        }}} \right) \div
	\exp \left(
-\frac{(c_\Omega - p_\Omega)^2}{2 \times 0.02^2}
\right) 
\label{reweight}
\end{equation} 
in order to
impose the correct penalty on the links. We ignore additional
constants that are independent of $\Omega_{DM} h^2$ in this expression
since the overall normalisation of the likelihood density is here
undetermined.  The re-weighting procedure necessarily degrades the
statistical spread of the results, however we find that the increase
in efficiency more than compensates for this effect. Below, we
re-weight different variables in order to investigate various features
in the results, but the method remains analogous to the one described
here.

\subsection{Bridge Sampling}

One of the numbers we will require from our MCMC samples is the ratio
of integrated posterior probabilities of $\mu>0$ ($P_+$) and $\mu<0$
($P_-$). This ratio will tell us the extent to which $\mu<0$ is
disfavoured over $\mu>0$.  Assuming a flat prior in the variables of
the model, the posterior probability is equal to the integrated
likelihood divided by a factor which does not depend upon model
hypotheses or parameters. Thus $P_-/P_+=\int d{\bf x} {\mathcal
L}_-({\bf x}) / \int d{\bf x}{\mathcal L}_+({\bf x})$, where
${\mathcal L}_{+,-}({\bf x})$ is the likelihood density of $\mu>0$
$(<0)$ mSUGRA respectively at parameter point $({\bf x})$.  One way to
estimate this ratio would be to include the sign of $\mu$ as a free
parameter in the Metropolis-Hastings procedure, to be chosen randomly
in a proposal point. This algorithm leads to large inefficiencies
because the $\mu>0$ and $\mu<0$ likelihood surfaces have a limited
overlap, meaning that too many proposals for an opposite sign of $\mu$
will be rejected. Also, the procedure would likely provide large
statistical fluctuations for the disfavoured sample, which we expect
to be the $\mu<0$ one. Even though it is disfavoured, we should like
to investigate its properties.

A simple way one might hope to evaluate the ratio ${P_-}/{P_+}$ is
\begin{equation}
\frac{P_-}{P_+} = \frac{1}{E_- \left[ \frac{{\mathcal L}_+}{{\mathcal L}_-}
    \right]} 
\approx \frac{1}{N} \sum_{t=1}^N \frac{{\mathcal L}_-
({\mathbf x_i^{(t)}})}{{\mathcal L}_+ ({\mathbf x_i^{(t)}})},
\end{equation}
where $E_-$ denotes the expectation with respect to the $\mu<0$
likelihood distribution \cite{radford}. $N$ denotes the number of MCMC
steps. 
Unfortunately, a simple importance sampling estimate of this kind does
not work if there are any valid points (${\mathcal L} \ne 0$) for one
sign of $\mu$ that are invalid (${\mathcal L} = 0$) for the opposite
sign of $\mu$.  In our case there are plenty of these dangerous
pairings, as sparticle mass or tachyonic bounds move around in
parameter space depending upon the sign of $\mu$.  To get around this
problem, we use a solution known as bridge sampling~\cite{radford}
with a ``geometric bridge''.  This allows us to generate a (biased)
estimator $r$ for the ratio ${P_-}/{P_+}$ as long as there is some viable
region of $\mu>0$ parameter space that is also viable for $\mu<0$.
The estimator for the ratio is constructed as follows:
\begin{equation}
 \frac{P_-}{P_+}  =
\frac{E_+ \left[ \sqrt{\frac{{\mathcal L}_-}{{\mathcal L}_+}}
  \right]}
{E_- \left[ \sqrt{\frac{{\mathcal L}_+}{{\mathcal L}_-}}
  \right]} \approx r \equiv
\frac{\sum_{t=1}^N \sqrt{\frac{{\mathcal L}_- ({\mathbf x}_+^{(t)})
}{{\mathcal L}_+ ({\mathbf x}_+^{(t)})}}}{\sum_{t=1}^N \sqrt{\frac{{\mathcal
	L}_+ ({\mathbf x}_-^{(t)}) 
}{{\mathcal L}_- ({\mathbf x}_-^{(t)})}}} , \label{bridge}
\end{equation}
where $(\mathbf x)^{(t)}_{+,-}$ are the parameter points of the links
in the $\mu>0$ or $\mu<0$ chains respectively.  Here, we have assumed
an equal number of links in each chain.  In summary, to calculate $r$,
we must run two chains, one for positive $\mu$ and one for negative
$\mu$, and for every link, record the likelihood one would have
obtained for identical input parameters except for the opposite sign
of $\mu$. 

\section{Convergence and Normalisation\label{sec:conv}}

\EPSFIGURE{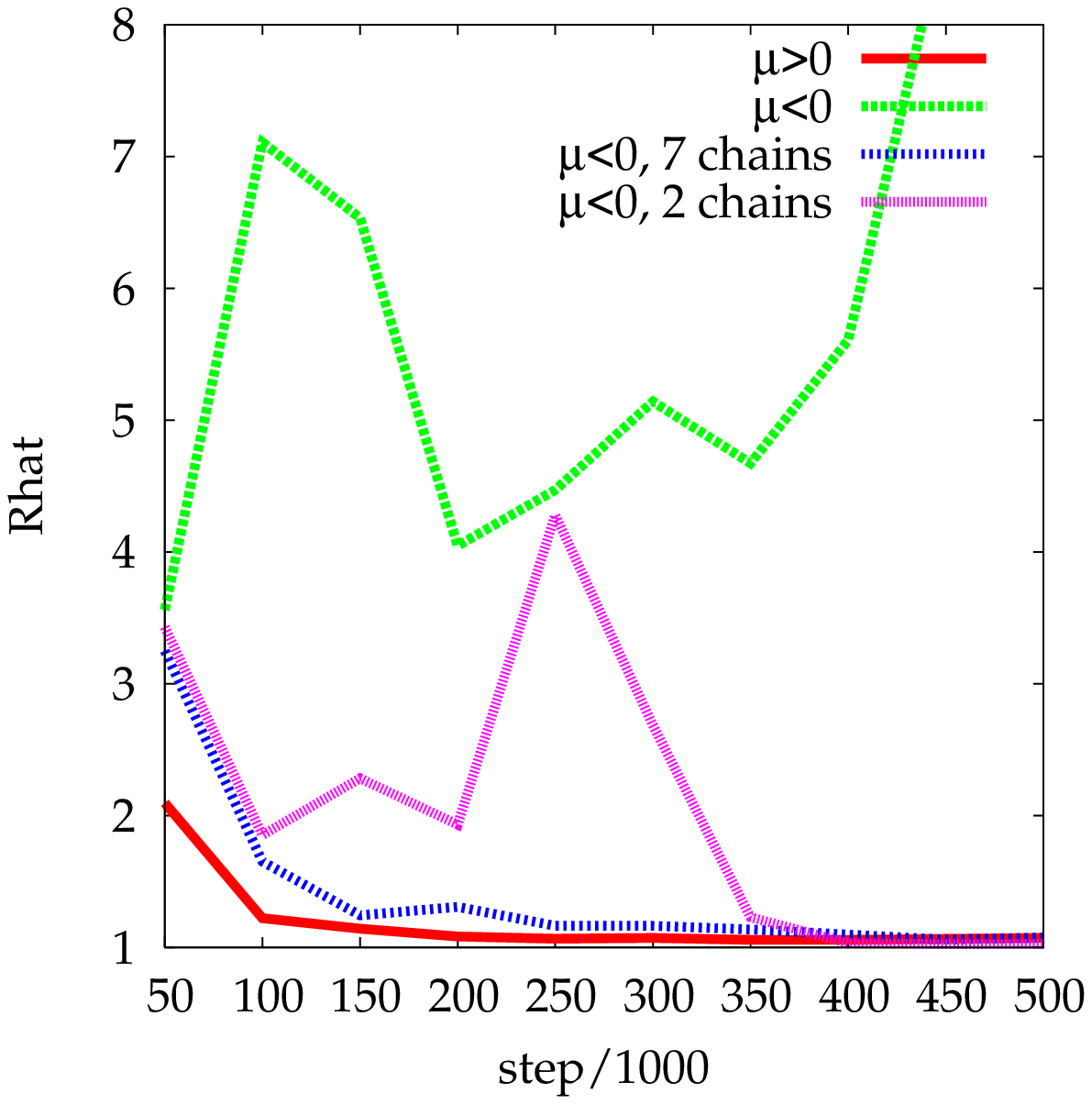,width=7cm}{Convergence statistics for the
  MCMCs. 
\label{fig:conv}}In order to evaluate the convergence of the MCMC chains, we always run 9
independent chains with different random starting points. By comparing the 
similarity of the resulting sampling densities of input parameters in
the chains, one can
construct~\cite{gelman} a measure of convergence $\hat R$. $\hat R$ is an upper
bound on the reduction in variance of parameters that would result from
running the chains for an infinite number of steps. The precise implementation
is  listed in Ref.~\cite{Allanach:2005kz}. Values close to 1 indicate
convergence of the chains.

\FIGURE{\twographst{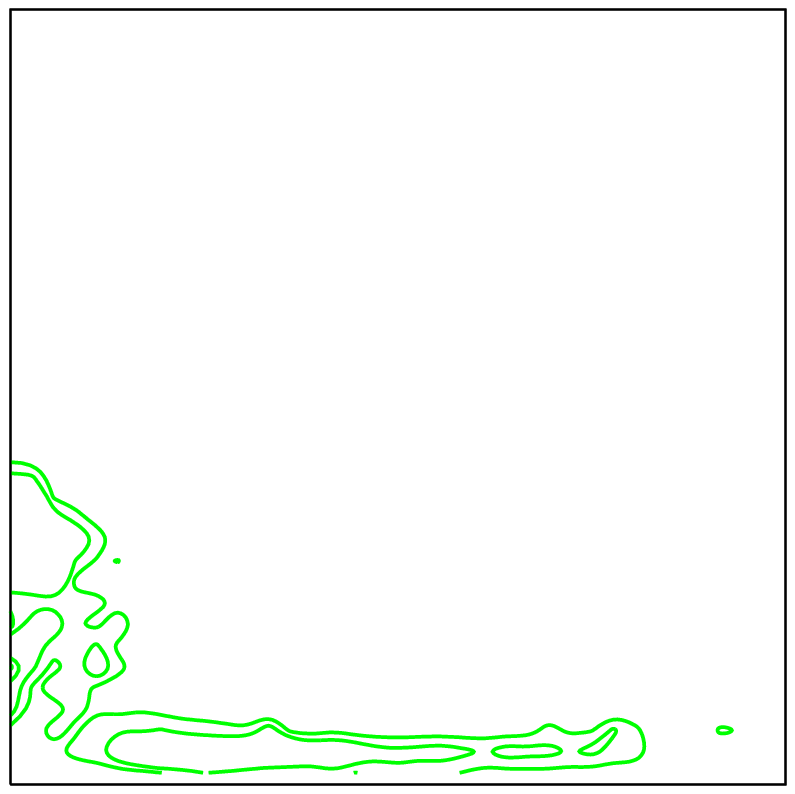}{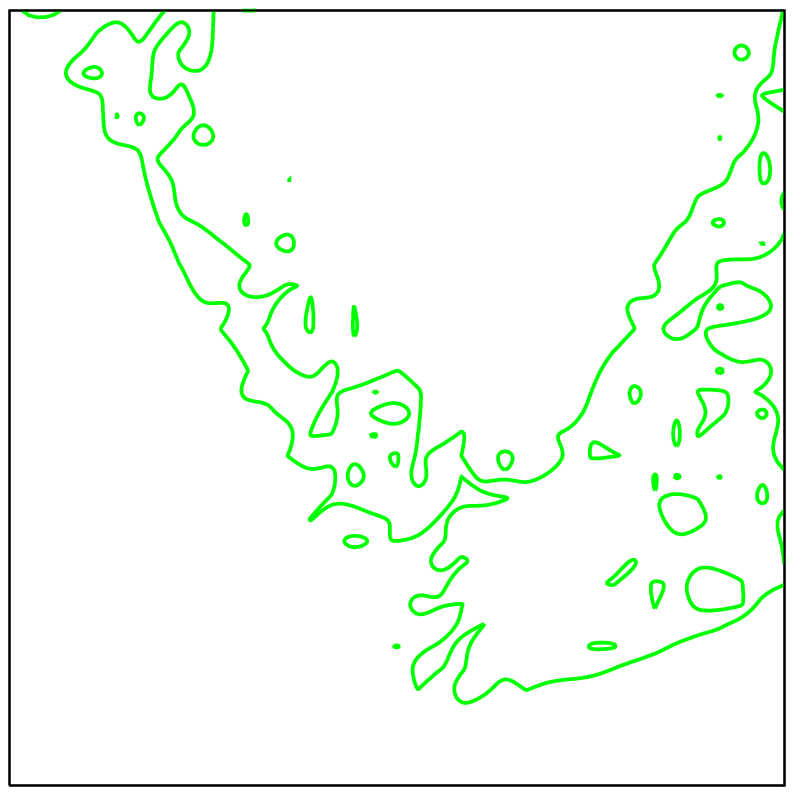}
\caption{The two types of negative $\mu$ samples: (a) the ``$\mu<0$, 2
  chains'' samples (later shown to be co-annihilation samples), and (b)
  the ``$\mu<0$, 7 chains'' samples (later shown to be
  non co-annihilation samples, dominated by resonant Higgs annihilation
  regions).  The posterior probability is indicated by the bar on the
  right hand side. The inner (outer) contours show the boundary of a
  $68\%$ $(95\%)$ confidence region.\label{fig:stop}}}
We run 9 chains of 500~000 points for $\mu>0$ mSUGRA and for the
$\mu<0$ dark side of mSUGRA\@.  The $\mu>0$ curve in
Fig.~\ref{fig:conv} shows good convergence is achieved by 500 000 MCMC
steps. However, the $\mu<0$ curve shows a problem: convergence is
never achieved. This is a serious difficulty as one could not draw any
quantitative statistical inferences from the non-converged chains.
Further inspection of the $\mu<0$ results shows that two of the
$\mu<0$ chains are in a completely different part of parameter space
than the other seven.  This indicates isolated maxima of likelihood
density which the MCMC has not been able to jump between in the finite
number of MCMC steps attempted\footnote{A proposal distribution with
longer tails, such as an $n-$dimensional Cauchy distribution, would
have more chance of making such a jump.}.  There is no balance between
the two isolated maxima in the sample.  Isolating the two anomalous
$\mu<0$ chains and calculating $\hat R$ between just them, we obtain
the ``$\mu<0$, 2 chains'' curve, which closely approaches 1 by 500~000
MCMC steps. Thus within this isolated maximum, convergence is
achieved.  The same can be said of the other ``$\mu<0$, 7 chains''
samples: they also converge amongst themselves. Thus, the shapes of
each isolated maximum are well determined, but the relative
normalisation of the two different types of negative $\mu$ samples is
not. 

\FIGURE{\onegraph{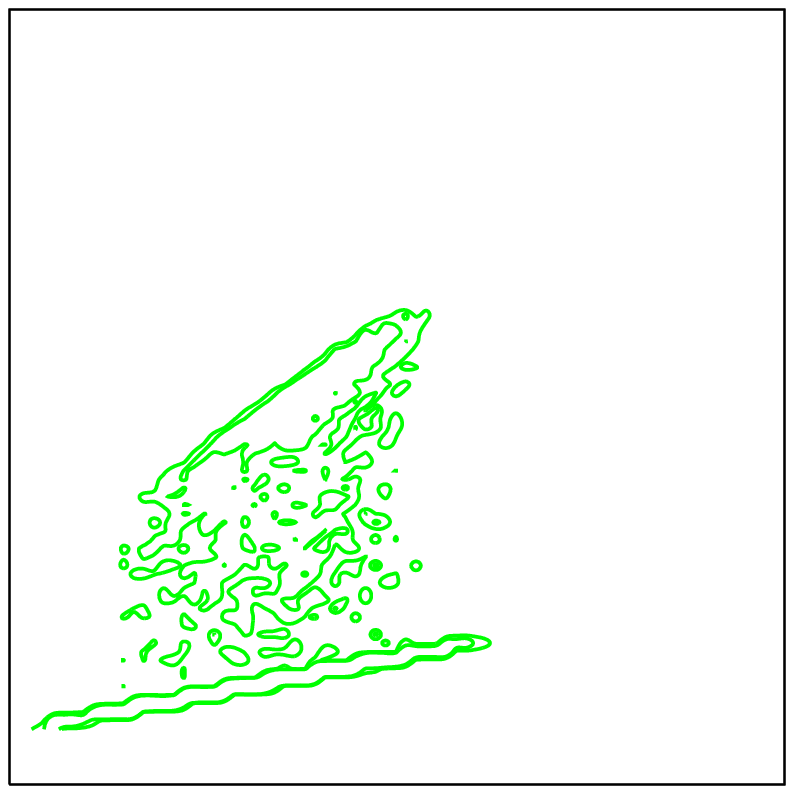}
\caption{Probability density in the lightest stop-lightest neutralino mass
  plane for the 2-chain sample. The posterior probability is indicated
  by the bar on the right hand side. The inner (outer) contours show
  the boundary of a $68\%$ $(95\%)$ confidence
  region. \label{fig:stop2}} }

In order to illustrate the two maxima, we marginalise the two types of
negative $\mu$ samples onto the $m_0-A_0$ plane in
Figs.~\ref{fig:stop}(a) and~\ref{fig:stop}(b).
  The maxima are isolated in this
plane (as well as in some other 2-parameter planes).  
The two
regions are completely separated. Their shape is primarily determined
by regions which efficiently deplete the relic density of neutralinos
which, in mSUGRA, is often higher than the WMAP3 constraint.  We
investigate the 2-chain sample in Fig.~\ref{fig:stop2}. 
In the figure,
there are two good-fit regions: where the stau
co-annihilates~\cite{Griest:1990kh} (at moderate values of $A_0$, higher
values of $m_{\tilde t}$ in the figure) with
the LSP ${\tilde \tau}_1 \chi_1^0 \rightarrow \tau \gamma$ and where
the lightest stop co-annihilates (at $A_0<-3$ TeV) with the LSP
${\tilde t} \chi_1^0 \rightarrow t g$ in the early universe, where
$m_{\chi_1^0} \approx m_{{\tilde t}_1}$, the lower
strip in the figure. There was no significant
stop co-annihilation~\cite{Boehm:9911,arnie,Ellis:2001nx} region for
$\mu>0$.  On the other hand, Fig.~\ref{fig:stop}(b) dominantly
consists of resonant Higgs annihilation
regions~\cite{Drees:1992am,Arnowitt:1993mg,Djouadi:2005dz}, where
$\chi_1^0\chi_1^0 \rightarrow h,A^0 \rightarrow b {\bar b}/\tau^+
\tau^-$ and the focus point region where the LSP has a significant
higgsino component and $\chi_1^0\chi_1^0 \rightarrow ZZ, WW, t
\bar{t}$~\cite{Feng:1999mn,Feng:1999zg,Feng:2000gh}.

We need a method to determine the relative normalisation of the 
2-chain co-annihilation sample and the 7-chain resonant higgs
annihilation sample.  Equivalently we need a method to determine the
ratio of the posterior probability ${\tilde P}_-^{\tilde t}$ of the 2-chain
co-annihilation sample and the posterior probability
${\tilde P}_-^{\slash\!\!\!{\tilde t}}$ of the 7-chain resonant higgs
annihilation sample. We use ${\tilde P}$ to denote the fact that the
posterior probabilities are un-normalised.
Fortunately, Eq.~\ref{bridge} provides us with a
solution: we first determine the normalisation of the $\mu>0$ sample
with respect to each separate $\mu<0$ sample, i.e.\ ${\tilde P}_-^{\tilde
t}/P_+$ and ${\tilde P}_-^{\slash\!\!\!{\tilde t}}/P_+$.  Since these
quantities individually have good convergence properties, their ratio is also
well determined:
\begin{equation}
\frac{{\tilde P}_-^{\slash\!\!\!{\tilde t}}}{{\tilde P}_-^{\tilde t}} = 
 \frac{{\tilde P}_-^{\slash\!\!\!{\tilde t}}}{P_+} \div \frac{{\tilde
 P}_-^{\tilde  t}}{P_+} 
=0.097 \div 0.063 = 1.53. \label{RofR}
\end{equation}
Normalising the probabilities as $a{\tilde P}_-^{{\tilde t}} \equiv
P_-^{{\tilde t}}$, $b{\tilde P}_-^{\slash\!\!\!{\tilde t}} \equiv
P_-^{\slash\!\!\!{\tilde t}}$, we fix $a$ and $b$ by imposing ${
  P}_-^{{\tilde t}} + {P}_-^{\slash\!\!\!{\tilde t}} = 1$ and
Eq.~\ref{RofR}. 
The relative posterior probabilities ratios are re-calculated
whenever alternative priors are investigated.
In section~\ref{sec:dark} where we present the total $\mu<0$
sample results, we present posterior probability densities with the correct
normalisation, according to this prescription.
The ratio of the probability of $\mu<0$ to $\mu>0$ is then determined simply
by: 
\begin{equation}
  \frac{P_-}{P_+} = \frac{P_-^{\tilde
 t}}{P_+} + \frac{P_-^{\slash\!\!\!{\tilde t}}}{P_+}. 
\end{equation} 

It is worth noting that, had we been unlucky, we might have obtained
only chains like those in the 7-chain sample. In that case we would
have carried on with the analysis without realising about the
different 2-chain sample, therefore any results achieved would have
been incomplete. An obvious question is: were there any other local
maxima that we have missed by not running enough chains?
Unfortunately, any fitting procedure is susceptible to this caveat and
there is no satisfactory answer. Finding a high but very narrow global
maximum is an unsolved problem in any number of dimensions.

\end{document}